\documentclass[manuscript]{acmart}
\usepackage[symbol]{footmisc}
\AtBeginDocument{%
  \providecommand\BibTeX{{%
    \normalfont B\kern-0.5em{\scshape i\kern-0.25em b}\kern-0.8em\TeX}}}

\usepackage{graphicx}
\usepackage{float}
\usepackage{subfigure}
\usepackage{listings}
\usepackage{booktabs}
\usepackage{tabularx}
\usepackage{colortbl}
\usepackage{color}

\begin{document}

\title{Agent-based Simulation for Online Mental Health Matching}


\author{Yuhan Liu}
\authornote{Both authors contributed equally to this research.}
\email{yl8744@cs.princeton.edu}
\affiliation{%
 \institution{Princeton University}
 \country{USA}
 }

\author{Anna Fang}
\authornotemark[1]
\email{amfang@andrew.cmu.edu}
\affiliation{%
 \institution{Carnegie Mellon University}
 \country{USA}
 }

\author{Glen Moriarty}
\affiliation{%
 \institution{7Cups}
 \country{USA}}

\author{Robert Kraut}
\email{robert.kraut@cmu.edu}
\affiliation{%
 \institution{Carnegie Mellon University}
 \country{USA}
 }

\author{Haiyi Zhu}
\email{haiyiz@cs.cmu.edu}
\affiliation{%
 \institution{Carnegie Mellon University}
 \country{USA}
 }
\renewcommand{\shortauthors}{Anonymized Authors}

\begin{abstract}
Online mental health communities (OMHCs) are an effective and accessible channel to give and receive social support for individuals with mental and emotional issues. However, a key challenge on these platforms is finding suitable partners to interact with given that mechanisms to match users are currently underdeveloped. In this paper, we collaborate with one of the world's largest OMHC to develop an agent-based simulation framework and explore the trade-offs in different matching algorithms. The simulation framework  allows us to compare current mechanisms and new algorithmic matching policies on the platform, and observe their differing effects on a variety of outcome metrics. Our findings include that usage of the deferred-acceptance algorithm can significantly better the experiences of support-seekers in one-on-one chats while maintaining low waiting time. We note key design considerations that agent-based modeling reveals in the OMHC context, including the potential benefits of algorithmic matching on marginalized communities.

\end{abstract}



\maketitle
\section{Introduction}
People are increasingly turning to online mental health communities (OMHCs) to obtain support for their mental and emotional health \cite{Stratford2019-nq, Turner1999-tj, Whittaker1997-jd}. OMHCs are a practical and accessible way for users to receive both informational and emotional support on a variety of mental and emotional concerns \cite{Naslund2016-gm}, with communities that offer either general support for any need or support for specified health issues. For example, communities such as 7Cups.com about any mental health issue through 1-on-1 peer counseling chats while platforms like BabyCenter.com provides targeted informational and emotional support resources for pregnant women \cite{gui2017investigating,baumel2016adaptation}. Support obtained through OMHCs has been found to be vital in maintaining and improving people’s well-being, such as reducing depression, developing meaningful relationships with others, and increasing trust in the efficacy of mental health treatment \cite{De_Choudhury2017-ox, Fan2014-db}. 

However, despite this ability of OMHCs to yield meaningful and positive relationships between users, they currently rely on naive methods (i.e. first-come-first-serve, solely based on topic of discussion) for members to find these relationships, without consideration of users’ unique characteristics and preferences\footnote{Some examples of these online mental health communities are 7cups.com, healthunlocked.com, 18percent.org, and betterhelp.com}. Prior work suggested that current first-come-first-serve-based matching systems do not adequately support users’ needs and capabilities; ineffective matching can also lead to fewer long-term relationships and reduced member commitment \cite{Fang_2022}. Moreover, this lack of purposeful matching methods may be particularly harmful for the experiences of marginalized communities, who have both strong preferences for mental health providers with a similar background to themselves and particular reliance on online communities for support \cite{Administration2001-gx, Augustaitis2021-fm, Bockting2013-rr, Kim2008-iq, Yadavaia2012-yg}. Given these challenges, there has been exploration of developing intelligent forms of matching to provide more optimal matches given users’ characteristics with minimal efforts on a user’s end \cite{Fang_2022}.

However, matching is a complicated mechanism design problem \cite{Abdulkadiroglu2005-wk}. It is challenging to meet all the possible matching goals between support seekers and providers, and prioritizing one goal might lead to worse outcomes in other goals. For example, there may exist trade-offs in prioritizing matching accuracy versus speed, or the support of vulnerable populations versus other groups. Given these challenges, tools such as agent-based simulation, which has long been used to apply social science theories to the design of HCI systems, may be useful for revealing the complexities and various trade-offs in matching protocols for online community designers \cite{Ren2014-pn, Ren2014-wb}. Importantly, running these low-cost, virtual experiments can predict community members' likely reaction to alternative design choices without disrupting the existing community dynamics. 

In this paper, we created a simulated “sandbox” based on agent-based modeling that allows community stakeholders to play with different matching algorithms and helps designers consider complex trade-offs in building new mechanisms for their community. In order to build this simulation based on a real OMHC, we collaborated with one of the world’s largest peer support platforms. We utilized this platform’s real dataset to ensure the simulation's key assumptions -- distributions of agents’ personal characteristics, patience levels, session lengths -- are identical to the distribution found from the real OMHC data. We also created two prediction models to predict a chat’s rating and whether a support-seeker and volunteer counselor pair will block one another. Our validation suggests that we replicated our study's research site’ current system accurately. We then used our simulation framework to experiment with novel algorithmic matching policies and utilized the prediction models to evaluate outcomes of these different policies. In particular, we created five algorithmic matching policies based on prior work studying OMHC users’ needs: \textbf{(1) first-come-first-serve} to minimizes wait time for support-seekers; \textbf{(2) similarity-based matching} that matches support-seekers and support providers together with similar personal characteristics like age and gender; \textbf{(3) rating-based matching} that optimize the predicted chat ratings, which is the main indicator of support-seekers’ satisfaction with a support provider in a chat; \textbf{(4) blocking-based matching} that minimizes the chance of blocking, which is designed based on users’ needs of avoiding the worst outcomes rather than aiming for the best conversations possible; \textbf{(5) rating-and-blocking-combined} which optimize support-seekers’ chat experiences while also avoiding the worst chat experiences, \textbf{(6) filter-based matching}, focusing on protecting vulnerable groups - namely, teenagers and gender minorities. We compared the five algorithmic matching policies, finding that different outcome goals are achieved through different algorithmic choices and . We also found that some goals can unexpectedly be achieved through optimizing certain outcomes, such as our finding that optimizing the chat experiences of users may be reached by targeting rating and blocking metrics, but this also may serve to protect vulnerable groups rather than strict filter-based matching.

In sum, we contribute a novel application of agent-based simulation to the online mental health context through creating a "sandbox" based on an existing active online mental health community. This simulation will also be contributed as an open-source project, such that OMHC designers and creators can freely explore and understand the trade-offs in designing real-world matching algorithms for their communities. We show the benefits of agent-based modeling for algorithm design in online communities by showcasing various trade-offs and design considerations revealed through experimenting with different matching policies, all without any disruption to the existing community. In addition to applying simulation to the OMHC context, we contribute practical findings for designing matching in OMHCs specifically, such as particular trade-offs in this context between algorithmic matching and waiting time of users, as well as how different community goals can be obtained through particular matching policies.

\section{Background and Related Work}

\subsection{Peer Support in Online Communities}
Professional help delivered through online interventions can benefit people experiencing a range of psychological difficulties, including significant improvements in depression and helping users to be more engaged with their mental health treatment \cite{doherty2012engagement}. The majority of online mental health support takes place in OMHCs where peers can speak anonymously about their experiences for free and 24/7, enabling people to share more intimate details of their health experiences without fear of judgment or stigma \cite{Prescott2020-kg}. Social support through online platforms has been shown to improve users’ well-being in numerous ways, such as reducing depression, lowering suicidal ideation, spreading information about mental health, and enabling help-seeking for stigmatized populations \cite{De_Choudhury2014-yc, Naslund2016-gm, Prescott2020-wy}. Additionally, users who provide support to others through OMHCs have been shown to benefit, including enhancing self-esteem and gaining better social abilities \cite{Bracke2008-pd, Salzer2013-gg}.

OMHCs are vital for groups that particularly struggle with stigma or challenges in access to mental health care. For example, adolescents often prefer online communities for help-seeking to speak about sensitive issues away from adult supervision \cite{Alvarez-Jimenez2016-uz, Birchwood2013-lg}. Online social support is crucial for LGBTQ+ youth in particular for maintaining mental health from the safety of their home while faced with difficulties such as isolation from family or friends \cite{Craig2014-yl, Fish2020-by, McConnell2017-tt}. Importantly, anonymity through these platforms can aid stigmatized and minority groups to engage in community-finding during intense challenges of stigma and poor mental health outcomes, and build close relationships without fear of prejudice \cite{Graham2014-ci,Cipolletta2017-by}. Some groups, such as females and gender minorities, also face unique challenges of sexual and verbal harassment while chatting on the platform \cite{Fang_2022}. Thus, special consideration for protection of vulnerable groups is crucial in building better matching mechanisms, given the unique benefits and challenges they face.

\subsection{Matching for Mental Health Purposes}

As people perceive those similar to themselves to be more trustworthy and likely to share their worldviews , past research has thoroughly explored how matching a client and therapists’ race, language, gender, and other variables impact therapeutic outcomes \cite{Felton1986-kf, Jackson1973-wo, Leong1986-iz, Sue1988-nz}. In traditional mental health resources, both racial and gender matching have been central focuses. Coleman et al. found through a meta-analysis of studies from 1970s to early 1990s that clients have strong preferences for choosing a therapist of the same race \cite{Coleman_undated-xf} and Felton et al. finding that clients consistently prefer therapists of their same gender \cite{Felton1986-kf}. However, studies have found that racial preferences do not yield significant effects on actual wellbeing outcome in therapy \cite{Maramba2002-ug} while gender matching results in higher likelihood to complete mental health treatment (particularly so regarding male clients with male therapists) \cite{Flaskerud1991-wo}. However, effects of racial matching may differ between racial groups; most notably, Black clients’ preference for Black therapists may also have beneficial mental health outcomes given mitigation of general mistrust towards mental health services and its association with White or European American values \cite{Maultsby1982-nm, Whaley2001-yi}. 

Work in online mental health services has also found that people prefer those similar to themselves. Fang and Zhu found that gender, age, and experience level are all significant factors in people’s preferences for online support relationships; in particular, gender matching for gender minorities and avoiding support-providers who were significantly younger resulted in support-seekers having a more positive experience \cite{Fang_2022}. Additionally, matching based on shared interests can lead towards more supportive relationships, although sharing mental health diagnoses was not necessarily a helpful component \cite{andalibi2021considerations}. Other aspects, such as matching users’ expectations of emotional or informational support, can also affect satisfaction \cite{Vlahovic2014-rp}.

\subsection{Matching for Other Purposes}
Although algorithmic matching has not been built in the OMHC context, algorithmic matching has been shown to have influential effects in other domains. Importantly, our study will utilize the applicant-proposing deferred-acceptance algorithm to match support-seekers and volunteer counselors on an online mental health community (elaborated on in Section 6.1), which has been used in numerous other contexts. For example, The National Resident Matching Program employs the applicant-proposing deferred-acceptance algorithm to produce a matching of new physicians to residency programs  \cite{Roth1999-zc}. This matching algorithm has also been extended to the educational context, such as the New York City Department of Education and Boston Public Schools, who match tens of thousands of entering students to public high schools \cite{Abdulkadiroglu2005-wk, Abdulkadiroglu2005-yn}. Other matching work done in education has focused on grouping students together for better educational outcomes, with conflicting results on how gender matching influences these outcomes \cite{Fenwick2001-qm, Stewart2018-xy}. Campbell and Campbell studied how racially matching students to mentors led to better long-term performance like higher GPA, graduation rate, and rate of entering graduate education \cite{Campbell2007-rz}. Additionally, matching for romantic and platonic relationships has been explored previously. Hitsch et al. found that users often have strong dating preferences for people similar to themselves across numerous dimensions with racial matching being the most influential \cite{Hitsch2010-nb}. Platonic friend finding in online social networks can also be optimized with algorithmic consideration of qualities like lifestyle and personality \cite{Bian2011-xq, Wang2015-th}. 

\subsection{Agent-based Modeling and Online Community Design}
Agent-based modeling, which is the simulation of the actions of agents to understand their behaviors and interactions under different conditions, has been applied to a wide range of research domains including economics, urban studies, and social sciences \cite{Huang2014-qj, Nigel_Gilbert2006-dx, Tesfatsion2006-ag}. Agent-based modeling has also been used to inform the design of online communities through simulating how different design choices impact desired outcomes, and has been used in past work to explore topics such as social influence and information propagation \cite{Alvarez-Galvez2016-oo, Delcea2017-vp, Du2017-jm, Van_Maanen2013-yt, Plikynas2015-gd, Ren2014-pn, Tan2011-bm}. Ren and Kraut have shown how agent-based modeling can be used to apply social science theories and understand trade-offs in design decisions, applying these methods to explore motivations for online community participation and how different moderation methods affect discussion in online communities \cite{Ren2010-jr, Ren2014-wb}. Sibley and Crooks found that utilizing agent-based models could be used to design recommendation algorithms to connect individuals and communities in online communities as they showed that networks with higher proportions of recommendation-based links produced more fragmented societies \cite{Sibley2020-cc}. Past work has also explored the intersection of emotions and online communities using agent-based frameworks, as studying emotion dynamics is different from other information or opinion propagation work given that emotions are short-lived, subjective, and require more sophisticated methods to extract. For example, Mitrovic and Tadic simulated the dynamics of the discussion-driven site Diggs to capture the emergence of emotional behaviors of its agents, while Schweitzer and Garcia explored how to observe collective emotion among an online community \cite{Mitrovic2012-hh, Schweitzer2010-pe}.

\section{Data}
In order to study algorithmic matching and simulation in the real online mental health context, we collaborated with one of the largest and currently active online peer support platforms.

\subsection{Research Site}
In order to study algorithmic matching and simulation in the real online mental health context, we collaborated with one of the largest existing peer support platforms that is currently an active and growing community. This online platform provides free 24/7 chat support, and has over 54 million members and 500,000 trained volunteer counselors. Users who sign up to seek support - who we will call “support-seekers” - can chat in 1-on-1 chat rooms with trained volunteers ("volunteer counselors").

The primary method of support on our study’s research site is through 1-on-1 chats between one support-seeker and one volunteer counselor. The current matching process is a self-selection by volunteer counselors where support-seekers send a request to join a live queue and wait to be picked by a volunteer counselor to begin a chat. Support-seekers also have the option to select one of many “topic tags” labeling their reason for seeking help (e.g. ‘depression’, ‘relationship stress’) but are not required to do so. No other information about support-seekers besides their wait time and possibly their topic tag is displayed in the queue. Note that support-seekers can cancel their chat request at any point, and may do so especially if waiting a long time without being selected by a volunteer counselor.

\subsection{Dataset}

The dataset consists of all chat messages between January 2020 to August 2020 on our study's research site, including the anonymized message text, timestamp, and user IDs involved, as well as logs for when a user blocked another user. The dataset also includes users’ signup dates and birth years. Note that no personally identifiable information about users is available. Chats can also be rated by support-seekers from 1 to 5 stars once the chat has continued for a certain length of time or after the chat ends; volunteer counselors are not able to rate a chat.\\ 

\noindent \textbf{Privacy, Ethics and Disclosure.} This paper used behavioral log data obtained through a collaboration with the online mental health community studied in this paper to conduct our analysis. All data was anonymized before analysis and no personally identifiable information was used in this study. Note that chat messages were only analyzed to find the gender distribution of users in order to generate agents for simulation purposes, which is described in Section 5.1.2. One author of this paper worked previously at an internship for this study’s research site. This work has been approved by the appropriate Institutional Review Board (IRB).

\subsection{Motivation}

Our work utilizes agent-based simulation to uncover the effects of alternative matching policies in the design of OMHCs. The design of our simulation is motivated by prior work that found OMHC users have considerable need for purposeful matching, and the potential for matching algorithms to lead to significant improvement for support-seeker experiences \cite{Fang_2022}. However, exploring different matching protocols has the potential to disrupt the particularly sensitive population of OMHC users through these new mechanisms' implementation and iteration processes. Given this, our work showcases how using agent-based simulation can reveal the impacts of using different characteristics considered important to users' experiences (i.e. gender, age, experience level) in algorithmic matching as well as reveal various algorithms' differing impacts between user groups. 

Prior work has found that gender, age, and experience level are important factors to giving and receiving support for OMHC users \cite{Fang_2022}. Users were found to consistently share their gender and age with one another when beginning a chat to manually find a suitable match. Gender identity was important to many users in finding suitable partnerships online, particularly for vulnerable communities (i.e. LGBTQ+, female users) who often sought out volunteer counselors similar to themselves for comfort and protection \cite{Fang_2022, Coleman_undated-xf}. Age was also an important factor for volunteer counselors; older support-seekers and younger volunteer counselors found it difficult to chat with one another given their differences in life experiences. Additionally, more experienced support-seekers also expressed growing frustrations and considerations of leaving the platform from matching with inexperienced volunteer counselors constantly. Behavioral log analysis complemented these reported experiences as there was measurable improvements to support-seekers' ratings of chats when including factors of gender, age, and experience level while matching users; for example, data analysis by \cite{Fang_2022} showed a striking chat rating improvement average of 1.18 stars out of 5 when nonbinary support-seekers chatted with a nonbinary volunteer counselor instead of a cisgender male volunteer counselor.

\section{Overview of Methods}
In order to showcase how agent-based modeling can be used to experiment with matching policies in the online mental health context, we followed three steps. 
\begin{enumerate}
    \item \textbf{Building agent-based simulation (Section 5).} We built a simulation to replicate the current system of our study’s research site. Specifically, we analyzed our study’s dataset to ensure the simulation's key assumptions -- distributions of agents’ personal characteristics, patience levels, session lengths -- are identical to the distribution found from the real OMHC data. We also developed two prediction models to predict a chat’s rating and whether a support-seeker and volunteer counselor pair will block one another.
    \item \textbf{Validating agent-based simulation (Section 6).} We validated our replication of our study’s research site by comparing the distribution of chat ratings, distribution of support-seeker and volunteer counselor pair who block one another, distribution of support-seekers’ waiting time before being matched, and overall successful matching rate between our simulation and the ground truth dataset of our study's research site. Our validation suggests that we replicated our study's research site’ current system accurately. 

    \item \textbf{Virtual Experiments (Section 7).} Lastly, we used our simulation framework as a “sandbox” to experiment with novel algorithmic matching policies and utilized the prediction models to evaluate the outcomes of these different policies. 
\end{enumerate}

\section{Building Agent-Based Simulation}
We first built a simulation based on agent-based modeling in order to replicate the current matching mechanisms of our study's research site, and utilized outcome prediction models to validate this replication.

\subsection{Model Assumptions}
We first detail the several assumptions made in order to build our simulation, all of which were implemented based on the real OMHC dataset. Specifically, we identify aspects such as the demographics of users, the number of support-seekers and volunteer counselors who come online at any given time, how long support-seekers wait for a chat, and the length of chat sessions to replicate this real OMHC behavior in our agent-based simulation.

\subsubsection{Simulation Period}
In our agent-based simulation, both support-seekers and volunteer counselors have the possibility of being matched during each “round”, or simulation period. We determined that a simulation period of one minute was fit for our model, as one minute is temporally granular enough to yield quick matching of users and derivable from our empirical data. 

\subsubsection{Generation of Agents}
Our simulation consisted of two types of agents: support-seekers and volunteer counselors. Each simulation period (i.e. each minute) generates new support-seeker and volunteer counselor agents that are eligible for matching. Agents are considered “online” (i.e. available to chat) immediately when they are generated. In order to determine the number of agents generated, we analyzed our study’s dataset over January to August 2020 to find the average number of online support-seekers and volunteer counselors for each simulation period over a week (i.e. 10,080 minutes). Our analysis is shown in Figure 2. The number of support-seekers always exceeds the number of volunteer counselors, with support-seekers online at any given minute ranging between 81 and 162 and volunteer counselors online at any given minute ranging between 72 and 161. The processes for agents to go “offline” (i.e. are deleted) are described in the following two sections. 

\begin{figure}[ht] 
\centering 
\includegraphics[width=0.6\textwidth]{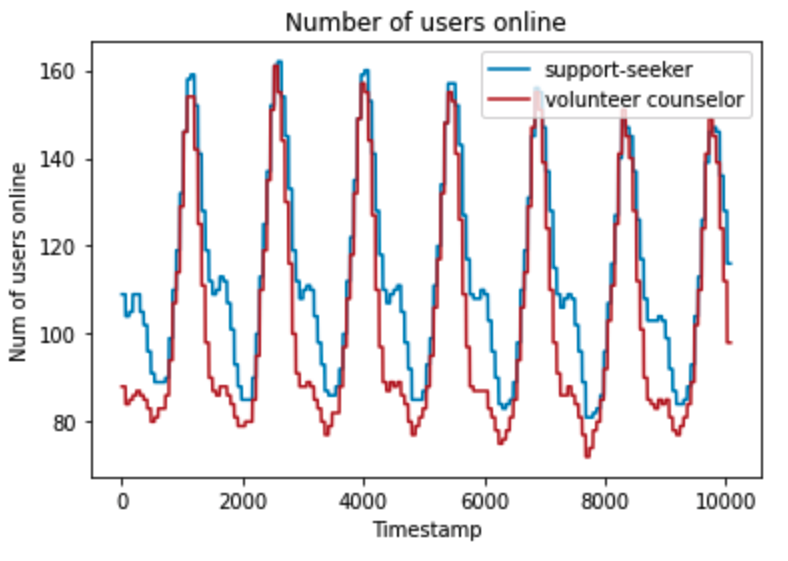} 
\caption{Number of online members and listeners in each simulation period. The number of online members has an average of 113.26 and standard deviation of 22.56. The number of online listeners has an average of 102.49 and standard deviation of 25.07.} 
\label{Fig:online_agents}
\end{figure}

\begin{figure}[ht]
\centering
\includegraphics[width=0.9\textwidth]{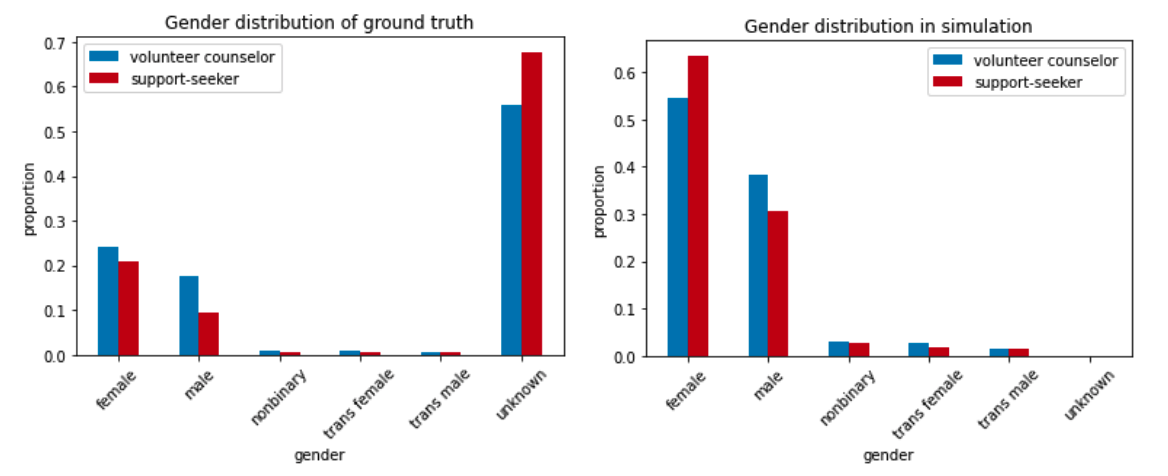}
\caption{Gender distribution in ground truth (left) vs. our simulation (right)}
\label{Fig.gender_distribution}
\end{figure}

When generated, each agent is also given several personal characteristics that may be significant to matching (gender, birth year, sign-up date, experience level) [4]. We analyzed the OMHC dataset to find the distribution of these characteristics at each simulation period, and assign agents’ characteristics so that the simulation’s distribution is identical to the distribution found from the real OMHC data. Although birth year, sign-up date, and experience level are all fields that are part of the raw dataset, we had to conduct a labeling process for users’ gender identity. Following prior work, we labeled gender according to whether a user had self-identified their own gender in chat logs (i.e. “I am a female”, “I am non-binary”), which had been found to be highly accurate and only mislabel gender for 0.8\% of OMHC users \cite{Fang_2022}. Using this process, we were able to label the gender of 35\% of support-seekers and 50\% of volunteer counselors in our dataset. We then applied the distribution of gender among those whose gender is known to our generated agents so that all agents have a gender characteristic. This distribution is shown in Figure \ref{Fig.gender_distribution}.

\subsubsection{Patience Level}
\begin{figure}[ht]
\centering
\includegraphics[width=\textwidth]{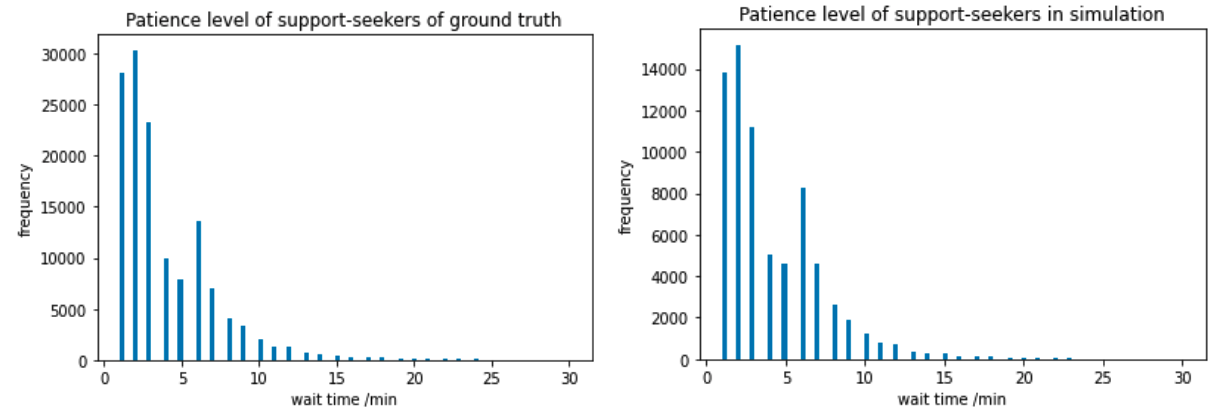}
\caption{Patience level of members in ground truth (left) versus our simulation (right). The mean of patience level in ground truth for our research site is 4.15 with standard deviation of 3.26 whereas our simulation's patience level has a mean of 4.16 with a standard deviation of 3.27. The Pearson correlation is 0.991.}
\label{Fig.patience}
\end{figure}

As support-seekers may leave the site while waiting for a chat, we also replicated the “patience level” of support-seekers. All support-seeker agents were given a number of minutes that they are willing to wait to be matched before they cancel their request (i.e. leave the platform). Support-seeker agents go offline when the number of simulation steps where the support-seeker remains unmatched exceeds their patience level. Similar to previous characteristics, we assigned patience levels to support-seeker agents according to the distribution we found from analyzing the dataset for how long support-seekers wait until canceling their chat requests in the queue. We assigned support-seeker agents a patience level according to the distribution of time for support-seekers to cancel their chat requests in the dataset. The patience levels of ground truth (in reality on our study’s research site) according to the log data is shown alongside the patience levels we assign to support-seekers in our simulation in Figure \ref{Fig.patience}, with a Pearson correlation of 0.991 between the two. 

\subsubsection{Chat Length}
\begin{figure}[ht]
\centering
\includegraphics[width=0.95\textwidth]{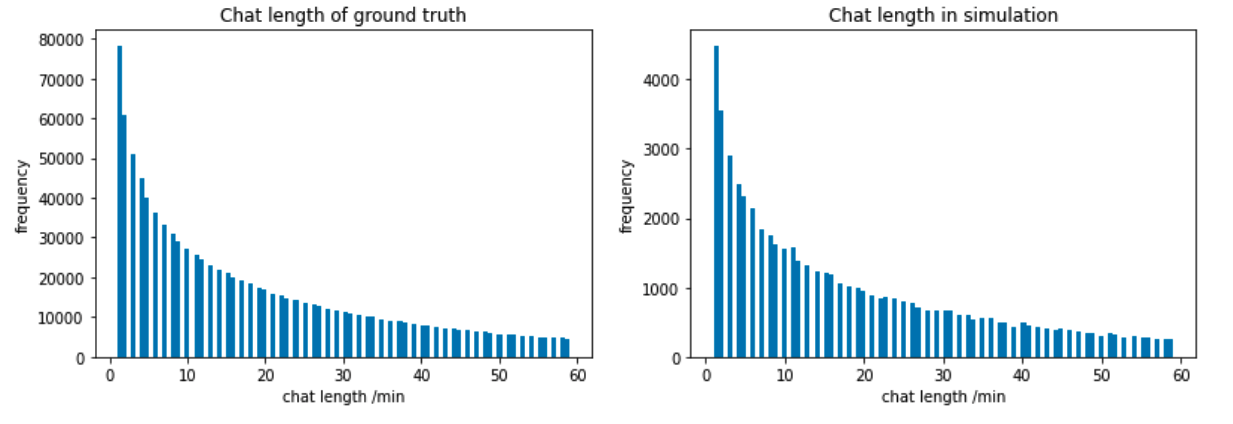}
\caption{Chat length of ground truth (left) versus our simulation (right). The mean of chat length in reality is 17.67 min with standard deviation of 15.44 min, whereas our simulation's chat length has a mean of 17.67 min with a standard deviation of 15.42 min. The Pearson correlation is 1.
}
\label{Fig.chatlength}
\end{figure}

Once a support-seeker and volunteer counselor are matched, their chat length is set in minutes and volunteer counselor agents go offline after a chat ends. We set the chat length in our simulation to follow the distribution of conversation length found in the log data. The distribution of chat length of ground truth found through log data versus our simulation's distribution is shown in Figure \ref{Fig.chatlength}, with a Pearson correlation of 1 between the two. 

\subsubsection{Matching Support-Seekers and Volunteer Counselors}
\begin{figure}[ht] 
\centering 
\includegraphics[width=0.6\textwidth]{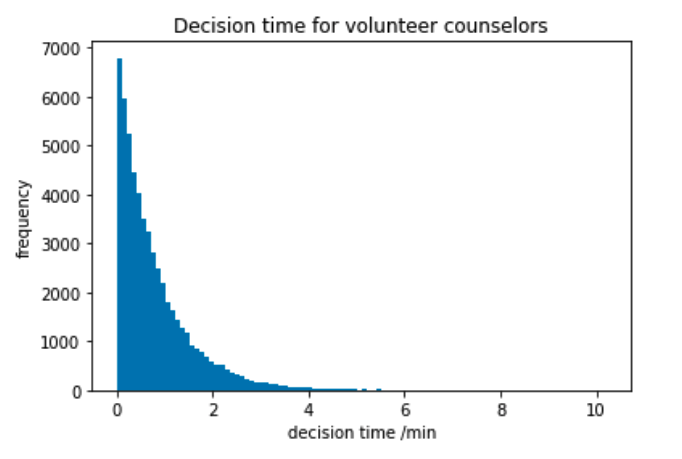} 
\caption{Distribution of decision time of volunteer counselors, which is modeled by an exponential distribution with lambda of 1.25.} 
\label{Fig:decision}
\end{figure}

Lastly, we describe the decision of actually matching a support-seeker and volunteer counselor together to chat. 
All agents who are online and not chatting in the current simulation period are considered available to be matched in the current round. In each simulation period, all volunteer counselor agents are presented with a list of all available support-seeker agents and pick a support-seeker to chat with depending on the matching policy. In the replication of the research site’s system, volunteer counselors pick a support-seeker to chat with randomly, following an exponential distribution model for the time it takes to make their choice (Figure \ref{Fig:decision}). 

\subsection{Outcome Prediction Models}
Although there are several outcome metrics that could be used to assess a support-seeker and volunteer counselor pair’s experience, the ability for a support-seeker to give the chat a rating of 1 to 5 stars is the most direct indicator of the match. Additionally, past work has found that users place more value on avoiding the worst chat experiences rather than aiming for the best given that extremely negative chats may include bullying and harassment and also may deter users from returning to the site \cite{Fang_2022}. Blocking another user is the main action that users take when they have a bad chat on the research site and is the most obvious reflection of a negative relationship between a support-seeker and volunteer counselor. Both support-seekers and volunteer counselors on our study’s OMHC can block each other through the chat interface and provide a reason for blocking, which includes categories like “Requesting/Sharing Personal Contact Info”, “Harassing/Threatening Behavior”, and “Inappropriate Sex Chat”. 

Note that, although our dataset includes good indicators of matching quality, there is no existing data that can be used to calculate matching quality for simulated results. Therefore, we created two outcome prediction models - a chat rating prediction model and a blocking prediction model. In the validation process, we used those two models to test whether our replicated simulation is an accurate representation of the current system. In Section 6, we use these prediction models to evaluate the effectiveness of our designed matching algorithms as well.

\subsubsection{Chat Rating Prediction}
We gathered all chat ratings in the dataset, with 80\% of samples from the dataset used as our training set while the remaining 20\% were used as our testing set. We balanced the training set using SMOTE oversampling, which generates artificial samples for minority classes based on existing samples using the K-nearest neighbor algorithm \cite{Chawla2002-hn} with results shown in Table \ref{Table:ratings_resample}. As independent variables in our chat rating prediction, we used inputs of both volunteer counselors' and support-seekers' gender, birth year, sign-up date, and experience level.  Our experiments included random forest, logistic regression, SVM, and decision tree models in order to find the best model to predict chat rating; results for each model’s accuracy and F1-score are shown in Table \ref{Table:ratings_algorithms}. Note that accuracy in this context means that the output must equal the true chat rating, and is considered incorrect if it outputs any of the other four chat ratings. Given that random forest outperformed all other models in accuracy and F1-score, we utilized the random forest classifier for our chat rating prediction model.

\begin{table}[ht]
	\centering
		\caption{Number of samples among each chat rating in the dataset, before and after SMOTE oversampling}

	\begin{tabular}{ccc}
		\toprule  
		Chat Rating(out of 5 starts)& \# of samples& \# of samples after SMOTE oversampling\\ 
		\midrule  
		1&2804&16730 \\
		2&799&16730 \\
		3&936&16730 \\
		4&2318&16730 \\
		5&16730&16730 \\
		\bottomrule  
	\end{tabular}

	\label{Table:ratings_resample}
\end{table}
\vspace{-\topsep}

\begin{table}[ht]
	\centering
		\caption{Accuracy and F1-score reports for all four machine learning algorithms.}

	\begin{tabular}{ccc}
		\toprule  
		ML algorithm& Accuracy& F1-score\\ 
		\midrule  
		Random Forest&0.55&0.58 \\
		Logistic Regression&0.45&0.5 \\
		SVM&0.49&0.53 \\
		Decision Tree&0.5&0.53 \\
		\bottomrule  
	\end{tabular}

	\label{Table:ratings_algorithms}
\end{table}

\subsubsection{Blocking Prediction}
In creating our training dataset for building the blocking prediction model, we labeled a support-seeker and volunteer counselor pair as 1 if at least one person blocked the other, and 0 otherwise. Similar to our chat rating prediction model, we utilized input fields of gender, birth year, sign-up date, and experience level for all agents as well as an 80\%/20\% split of samples from the dataset for training and test, respectively. Table \ref{Table:blocking_resample} shows the number of samples in our training set, along with results from oversampling to balance the training set. We proceeded again with random forest classification as it resulted in the best performance with highest precision and recall scores (shown in Table \ref{Table:blocking_algorithms}). 

\begin{table}[ht]
	\centering
		\caption{Number of samples in the dataset before and after SMOTE oversampling}

	\begin{tabular}{ccc}
		\toprule  
		Blocked/Not-Blocked Status& \# of samples& \# of samples after SMOTE oversampling\\ 
		\midrule  
		blocked&2804&16370 \\
		not-blocked&799&16730 \\
		\bottomrule  
	\end{tabular}
	
	\label{Table:blocking_resample}
\end{table}

\begin{table}[ht]
	\centering
		\caption{Accuracy and F1-score report for all four machine learning algorithms}

	\begin{tabular}{ccc}
		\toprule  
		ML algorithm& Accuracy& F1-score\\ 
		\midrule  
		Random Forest&0.89&0.9 \\
		Logistic Regression&0.7&0.78 \\
		SVM&0.84&0.87 \\
		Decision Tree&0.71&0.78 \\
		\bottomrule  
	\end{tabular}

	\label{Table:blocking_algorithms}
\end{table}

\section{Validation of Agent-Based Simulation}
After building our simulation model, we validated its accuracy in replicating the research site's system in four aspects detailed below.

\subsection{Number of Users Online}
\begin{figure}[ht] 
\centering 
\includegraphics[width=0.9\textwidth]{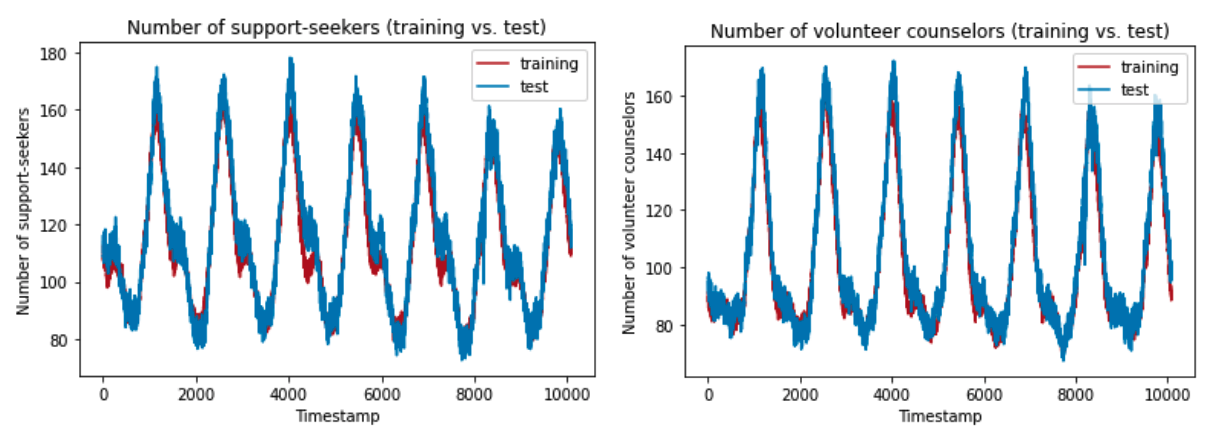} 
\caption{Distribution of number of support-seekers (left) and volunteer counselors (right) in training and test set. Pearson correlation of number of support-seekers between training and test set is 0.974 while Pearson correlation of the number of volunteer counselors between training and test set is 0.982. } 
\label{Fig:users_online}
\end{figure}

To validate replication of OMHC system, we split the dataset to a 6-month training set and 2-month test set. Figure \ref{Fig:users_online} compares the number of online support-seekers and volunteer counselors, respectively, in each minute between training set and test set in a week’s period of time. We found similar distributions between the training and test set, with Pearson correlations of 0.974 and 0.982 respectively. 

\subsection{Chat Ratings}
We utilized our chat rating prediction model to compare the distribution of chat ratings between our simulation and ground truth found through analysis of the dataset. We found that we accurately simulated rating distributions on the research site, with a Pearson correlation of 0.984 between our replication and ground truth. Our replication's proportions of ratings from 1-star to 5-stars were, respectively, 12.21\%, 4.08\%, 5.53\%, 9.32\%, 68.85\%, while ground truth proportions were 15.18\%, 3.51\%, 4.56\%, 10.63\%, and 66.12\% as shown in Figure \ref{Fig:rating_blocking_compare}.

\subsection{Blocking}
\begin{figure}[ht] 
\centering 
\includegraphics[width=0.9\textwidth]{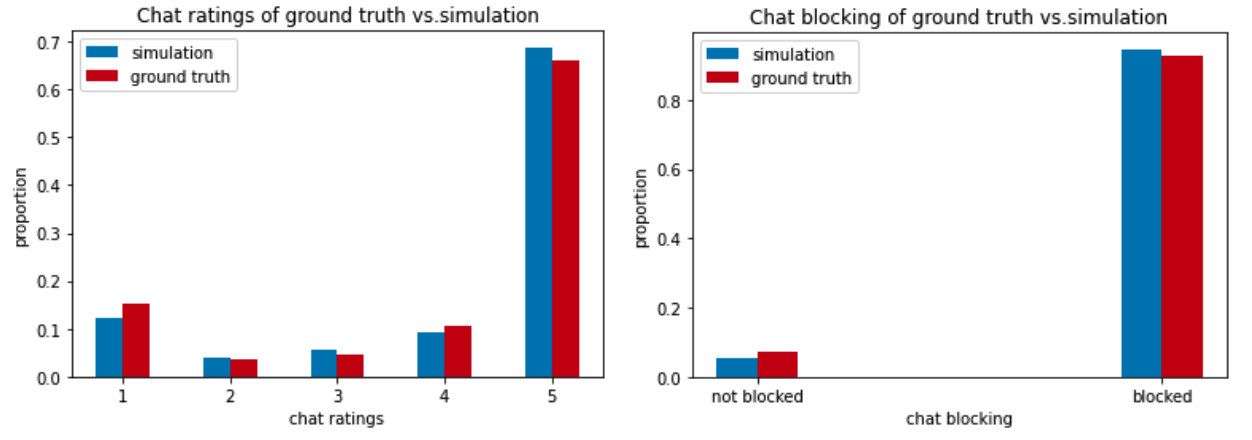} 
\caption{Comparison of blocking between reality versus our simulation.} 
\label{Fig:rating_blocking_compare}
\end{figure}

Similarly, we validated our replicated system using the blocking prediction model. On the research site, the proportion of blocked support-seeker and volunteer counselor pairs is 5.3\%, compared to our simulation’s proportion of 7.37\%. Utilizing the blocking prediction model, we found similar distributions of pairs that engage in blocking as in ground truth with a Pearson Correlation of 0.963 as shown in Figure \ref{Fig:rating_blocking_compare}, leading to further confidence that our simulation is an accurate representation of the current platform’s system.

\subsection{Waiting Time for Matched Pairs}
\begin{figure}[ht]
\centering
\includegraphics[width=0.9\textwidth]{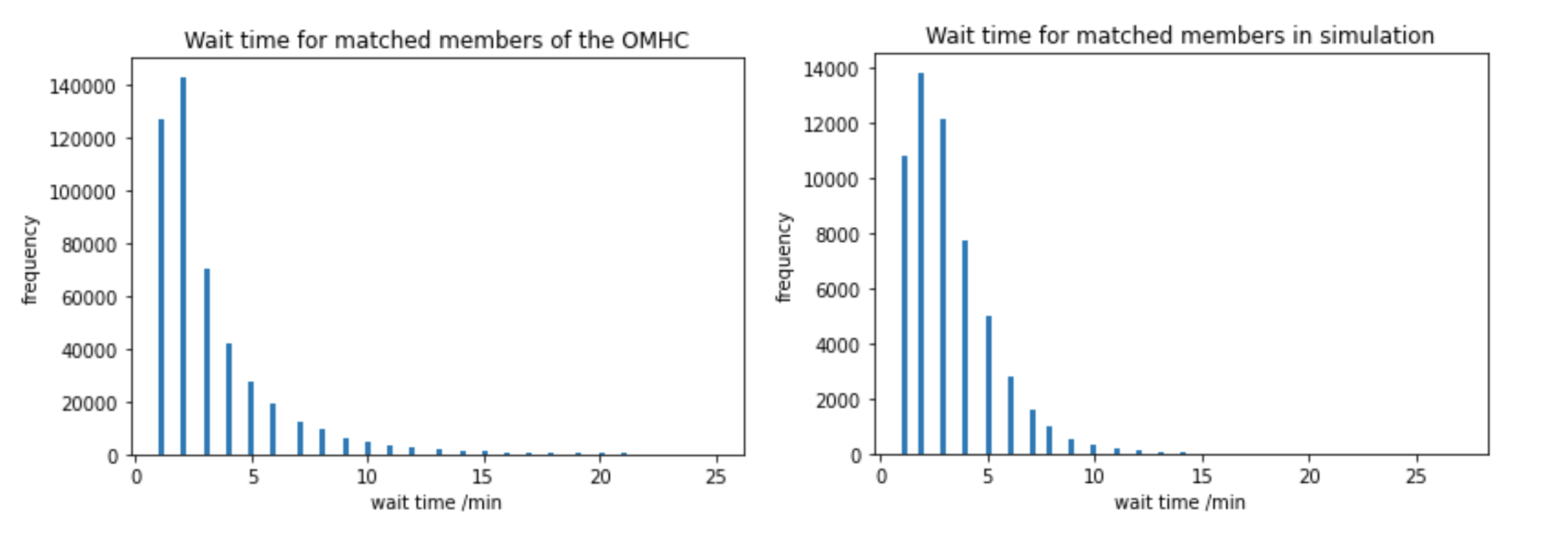}
\caption{Waiting time in reality (left) vs. simulation (right). The mean of waiting time in reality is 3.2 min with standard deviation of 2.9 min, whereas our simulation’s chat length has a mean of 3.2 with a standard deviation of 2.02 min. The Pearson correlation is 0.995.  }
\label{Fig.waiting_time_compare}
\end{figure}

We also validated using the waiting time of support-seekers matched in our simulation. We found that the distribution of waiting time for support-seekers who are matched with a volunteer counselor is similar to the distribution in the real online community system, although with a lower standard deviation, as shown in Figure \ref{Fig.waiting_time_compare}. The Pearson correlation of waiting time between reality and our simulation is 0.955. 

\subsection{Matching Rate}
We also compared the proportion of support-seekers who were matched with some volunteer counselor to the total number of support-seekers available to be matched. Note that support-seekers may cancel their chat request if they are unmatched for longer than their patience level. In order to evaluate ground truth, we analyzed the proportion of chats taken by volunteer counselors on the research site to the total number of requests by support-seekers in the the research site queue. We found that our simulation resulted in an overall 78.35\% matching rate in a week while ground truth showed an average of 83.27\% matching rate across all weeks in our dataset. Although our simulation’s matching rate is slightly lower than ground truth, the likely explanation is that our simulation framework only allows support-seekers and volunteer counselors to engage in one chat at a time whereas the research site allows for volunteer counselors to take multiple chats at once if they desire. Given this, we found that our simulation’s matching rate is acceptable and still closely resembled the actual state of the research site.

\section{Virtual Experiments}
Next, we utilized our simulation as a “sandbox” to test different matching algorithms. Below, we review the matching algorithms in our experiment and their outcomes on metrics of chat ratings, blocking, waiting time, and matching rates.

\subsection{Applicant-Proposing Deferred-Acceptance Algorithm}
Our research problem consisted of two types of agents (support-seekers and volunteer counselors) with preferences and personal characteristics for matching. As a result, we consider it akin to the stable marriage problem, which seeks to find a stable matching between two classes of elements where both sides have an ordering of preferences. Similarly, we sought a stable matching assignment between support-seekers and volunteer counselors such that no pair of support-seekers and volunteer counselors both prefer one another over their current match. Thus, we employed the \textbf{applicant-proposing deferred-acceptance algorithm} as our matching algorithm \cite{Abdulkadiroglu2005-wk, Gale1961-ig, Roth1999-zc}.

We briefly describe the applicant-proposing deferred-acceptance algorithm in our research context below, adapted from the established matching algorithm for New York Public Schools \cite{Abdulkadiroglu2005-wk}.\\

In each simulation period, there are two sets of agents: \textbf{support-seekers} $S=\{s_1,\, s_2, \,...,\,s_n\}$ and \textbf{volunteer counselors} $V=\{v_1, \,v_2, \,...,\, v_m\}$. Each support-seeker has an ordered \textbf{preference list} $P(s)=\{v_1,\, v_2,\, .., \,v_m\}$ where their first choice is volunteer counselor $v_1$, second choice is $v_2$, etc.\footnote{Note that “apply”, “hold” and “reject” are used to describe the algorithm’s actions, not the human users’ actions. }
\begin{itemize}
\item[1. ]Each support-seeker “applies” to her highest ranked counselor according to their preferences, and each volunteer counselor “holds” her highest ranked application and rejects the rest.

\item[2. ]At any stage at which a support-seeker has been rejected, she “applies” to her next most preferred volunteer counselor who the support-seeker agent prefers (if one remains). Each volunteer counselor holds its most preferred set of applications and “rejects” the rest. 

\item[3. ]The algorithm stops when no rejections are issued, and each volunteer counselor is matched to the applicants she is holding.
\end{itemize}
Any agent who is not matched in this simulation period is marked as “waiting”, and continues to be available for matching in the next matching period (along with any newly generated agents). Appendix A shows the pseudocode for the above algorithm.

\subsection{Matching Algorithms}
The key design choice in algorithmic matching is to construct the preference lists of support-seekers and volunteer counselors. Unlike past work done in public school matching and physician matching that involve a handful of school/hospital options, there are hundreds of volunteer counselors available for any support-seeker at any given time. Since it is impossible to ask each support-seeker to provide preferences for each individual volunteer counselor (and vice versa), there is a need for support-seeker and volunteer counselor preferences to be generated using rules or prediction models. The six algorithms are described below. We implemented different methods to generate support-seekers’ and volunteer counselors’ preferences, all based on our prior understanding of the research site support-seekers’ needs. Each algorithm varies in how it “recommends” a support-seeker to each volunteer counselor in each simulation period - volunteer counselors have a 90\% chance of taking the recommendation and a 10\% chance of randomly picking another available support-seeker. 

\begin{itemize}
\item \textbf{First-Come-First-Serve}: Support-seekers are ranked in volunteer counselors’ preference lists by decreasing waiting time. The goal of first-come-first-serve is to improve the number of successful matches and minimize the number of support-seekers who quit from exceeding their patience level. 

\item \textbf{Similarity-Based}: As past work has found that volunteer counselors in OMHCs find it difficult to help support-seekers with large age gaps from themselves and there may be benefit to matching vulnerable gender groups with one another \cite{Fang_2022}, we match support-seekers and volunteer counselors together with similar personal characteristics. We define a vector for each agent with their birth year, gender, signup date and experience level and calculate two agents’ similarity using the cosine similarity between their vectors. Agents with higher similarity get ranked higher in other agents’ preference lists.

\item \textbf{Rating-based}: Support-seekers can directly indicate their experience with a volunteer counselor in a chat by using the chat rating system, and thus we designed a rating-based algorithm where each agent’s preference list is created by our chat rating prediction model. The predicted rating for a potential support-seeker and volunteer counselor pair (s,v) is represented as  $R_{(s,v)}\in\{1,\,2,\,3,\,4,\,5\}$ and an agent’s preference list is ordered from the agents with the highest predicted rating to the agents with the lowest predicted rating. 

\item \textbf{Blocking-based}: Given that OMHC users generally feel that the primary objective of algorithmic matching should be to avoid the worst outcomes rather than aiming for the best conversations possible \cite{Fang_2022}, we create a blocking-based algorithm where agents’ preference lists are based on our blocking predictor. The blocking prediction result for a potential support-seeker and volunteer counselor pair $(s,v)$ is represented as $B_{(s,v)}\in\{0,1\} $ with 1 being prediction of either one of the agents blocking the other and 0 being prediction of neither one of the agents blocking the other. All members who are predicted as 1 by the blocking predictor are put at the bottom of listeners’ preference lists; the rest of the preference list is then ranked based on the members’ waiting time. 

\item \textbf{Rating-and-Blocking-Combined}: In order to optimize support-seekers’ chat experiences while also avoiding the worst chat experiences, we implemented a Rating-and-Blocking-Combined algorithm that never matches a support-seeker and volunteer counselor together if they are predicted to engage in blocking but otherwise matches support-seekers and volunteer counselors who are predicted to have a higher chat rating. As such, we combine rating and blocking to create a matching “score” C that is defined as: 

$$ C_{(s,v)}=
\begin{cases}
-1,  & \text{if }B_{(s,v)}=1 \\
R_{(s,v)} & \text{otherwise }
\end{cases}
$$
\\

\item \textbf{Filter-based}: Suggested by stakeholders in our study's research site, we implement filter-based methods in order to protect vulnerable groups - namely, teenagers and gender minorities. Agents considered to be teenagers are born after 2002, and agents considered part of gender minority groups are not cisgender female nor cisgender male. The filter-based method includes three different pools for agents: teenager pool, gender-minority pool, and all others. volunteer counselors can only select support-seekers who are in the same pool as themselves. Within each pool, volunteer counselors pick up support-seekers randomly. 
\end{itemize}

\subsection{Results}
We review the outcomes of experimenting with the six algorithms above using our simulation sandbox, shown in Table \ref{Table:simulated_result}. The outcome metrics used to evaluate the algorithms are defined as:

\vspace{-\topsep}
\begin{itemize}
\item \textbf{\% of high ratings}: Percentage of simulated chats that resulted in a chat rating at least 4 out of 5 stars from our chat rating prediction model
\item \textbf{\% of low ratings}: Percentage of simulated chats that resulted in a chat rating of under 3 out of 5 stars from our chat rating prediction model
\item \textbf{Average rating}: Average rating for all pairs in simulation predicted by the rating prediction model.  
\item \textbf{\% of blocked pairs}: Percentage of pairs in the simulation that are predicted as “block”
\item \textbf{Average waiting time (matched)}: average waiting time of support-seekers before they match in the simulation
\item \textbf{Average waiting time (unmatched)}: average waiting time of support-seekers whose waiting time exceeds patience level and quit
\item \textbf{Matching rate}: the ratio of support-seekers who were successfully matched (i.e. chatted with a volunteer counselor) to all support-seekers in the simulation
\end{itemize}

\vspace{-\topsep}

\definecolor{darkred}{RGB}{221,126,107}
\definecolor{lightred}{RGB}{230,184,175}
\definecolor{darkgreen}{RGB}{147,196,125}
\definecolor{lightgreen}{RGB}{217,234,211}
\begin{table}[ht]
	\centering
		\caption{Outcome metric results for different algorithms’ performance. For each outcome, the worst result is highlighted in dark red and the second worst result is highlighted in light red. The best result is highlighted in dark green and the second best result is highlighted in light green. (*) We still note the rating and \% blocked pairs outcome metrics even for rating- and blocking-based algorithms, respectively, although these outcome predictors (Section 5.2) were used in the algorithms themselves.}

	 \resizebox{\textwidth}{!}{
	\begin{tabularx}{\textwidth}{lX<{\centering}X<{\centering}X<{\centering}X<{\centering}X<{\centering}X<{\centering}X}
		\toprule  
		&\% of high rating& \% of low rating& Average rating& \% of blocked pairs& Avg waiting time (matched)&  Avg waiting time (unmatched)& Matching rate\\ 
		\midrule  
		Replication of ground truth (research site)&78.18\%&16.29\%&4.19&5.3\%&\cellcolor{darkgreen}3.19&3.7&\cellcolor{lightred}78.35\% \\
	    
	    First-Come-First-Serve&78.06\%&16.38\%&4.18&\cellcolor{lightred}6.49\%&\cellcolor{darkred}3.68&\cellcolor{darkgreen}2.73&\cellcolor{lightgreen}81.83\% \\
	    
	    Similarity-based&\cellcolor{lightred}77.61\%&\cellcolor{lightred}17.19\%&\cellcolor{lightred}4.15&\cellcolor{darkred}6.49\%&\cellcolor{darkgreen}3.16&\cellcolor{lightred}3.82&79.09\% \\
	    
	    Rating-based&\cellcolor{darkgreen}88.31\%*&\cellcolor{darkgreen}4.58\%*&\cellcolor{darkgreen}4.58*&3.84\%&3.58&3.04&80.45\% \\
	    
	    Blocking-based&88.22\%&9.67\%&4.21&\cellcolor{darkgreen}2.86\%*&\cellcolor{lightred}3.66&\cellcolor{lightgreen}2.75&\cellcolor{darkgreen}82.45\% \\
	    
	    Rating-blocking-combined&\cellcolor{lightgreen}88.24\%*&\cellcolor{lightgreen}8.56\%*&\cellcolor{lightgreen}4.56*&\cellcolor{lightgreen}2.95\%*&3.52&3.05&80.93\% \\
	    
	    Filter-based&\cellcolor{darkred}76.99\%&\cellcolor{darkred}17.32\%&\cellcolor{darkred}4.14&6.01\%&3.24&\cellcolor{darkred}3.86&\cellcolor{darkred}71.62\% \\
		\bottomrule  
	\end{tabularx}}
	
	\label{Table:simulated_result}
\end{table}

\begin{table}
\centering
\caption{Outcome metric results for different algorithms’ performance broken down by agents who are (1) non-teenagers (2) teenagers (3) non gender minorities (4) gender minorities. Bolded are particular results of note that are reviewed in Section 7.3 Results.}
\label{Table:broken_down_simulated_result}
\begin{tabular}{llllllll} 
\toprule
                                                                                                                                                                             & \begin{tabular}[c]{@{}l@{}}\textbf{\% of }\\\textbf{high}\\\textbf{rating}\end{tabular}   & \begin{tabular}[c]{@{}l@{}}\textbf{\% of }\\\textbf{low}\\\textbf{rating}\end{tabular}  & \begin{tabular}[c]{@{}l@{}}\textbf{Avg}\\\textbf{rating}\end{tabular}                                                 & \begin{tabular}[c]{@{}l@{}}\textbf{\% of}\\\textbf{blocked}\\\textbf{pairs}\end{tabular}                                                & \begin{tabular}[c]{@{}l@{}}\textbf{Avg }\\\textbf{waiting time}\\\textbf{(matched)}\end{tabular} & \begin{tabular}[c]{@{}l@{}}\textbf{Avg }\\\textbf{waiting time}\\\textbf{(matched)}\end{tabular} & \begin{tabular}[c]{@{}l@{}}\textbf{Matching}\\\textbf{rate}\end{tabular}                                                       \\ 
\midrule
\begin{tabular}[c]{@{}l@{}}\textbf{Replication}\\\textbf{ground truth}~ ~ \\~ ~ Non-Teenagers\\~ ~ Teenagers\\~ ~ Non-Gender Minorities\\~ ~ Gender Minorities\end{tabular}        & \begin{tabular}[c]{@{}l@{}}\\~\\76.09\%\\87.94\%\\77.47\%\\89.27\%\end{tabular} & \begin{tabular}[c]{@{}l@{}}\\~\\17.72\%\\9.58\%\\16.74\%\\9.23\%\end{tabular} & \begin{tabular}[c]{@{}l@{}}\\~\\4.12\\4.51\\4.16\\4.6\end{tabular}  & \begin{tabular}[c]{@{}l@{}}\\~\\6.29\%\\0.66\%\\5.53\%\\1.76\%\end{tabular} & \begin{tabular}[c]{@{}l@{}}\\~\\3.19\\3.23\\3.2\\3.15\end{tabular}                     & \begin{tabular}[c]{@{}l@{}}\\~\\3.69\\3.78\\3.7\\3.75\end{tabular}                       & \begin{tabular}[c]{@{}l@{}}\\~\\78.61\%\\77.84\%\\78.48\%\\78.3\%\end{tabular}  \\
\begin{tabular}[c]{@{}l@{}}\textbf{First-Come-}\\\textbf{First-Serve}~ ~\\~ ~ Non-Teenagers~ ~ \\~ ~Teenagers~ ~ \\~ ~Non-Gender Minorities~ ~ \\~ ~Gender Minorities\end{tabular} & \begin{tabular}[c]{@{}l@{}}\\~\\75.9\%\\88.08\%\\77.32\%\\89.60\%\end{tabular}  & \begin{tabular}[c]{@{}l@{}}\\~\\17.86\%\\9.54\%\\16.85\%\\8.96\%\end{tabular} & \begin{tabular}[c]{@{}l@{}}\\~\\4.1\\4.52\\4.16\\4.61\end{tabular}  & \begin{tabular}[c]{@{}l@{}}\\~\\6.43\%\\0.58\%\\5.66\%\\1.27\%\end{tabular} & \begin{tabular}[c]{@{}l@{}}\\~\\3.68\\3.69\\3.68\\3.67\end{tabular}                    & \begin{tabular}[c]{@{}l@{}}\\~\\2.73\\2.74\\2.74\\2.7\end{tabular}                       & \begin{tabular}[c]{@{}l@{}}\\~\\81.8\%\\81.98\%\\81.88\%\\81.05\%\end{tabular}  \\
\begin{tabular}[c]{@{}l@{}}\textbf{Similarity-based}~ \\~ ~ Non-Teenagers~ ~ \\~ ~Teenagers~ ~ \\~ ~Non-Gender Minorities~ ~ \\~ ~Gender Minorities\end{tabular}                   & \begin{tabular}[c]{@{}l@{}}\\75.22\%\\89.26\%\\76.86\%\\89.48\%\end{tabular}    & \begin{tabular}[c]{@{}l@{}}\\19\%\\8.38\%\\17.69\%\\9.27\%\end{tabular}       & \begin{tabular}[c]{@{}l@{}}\\4.07\\4.57\\4.12\\4.6\end{tabular}     & \begin{tabular}[c]{@{}l@{}}\\7.69\%\\0.88\%\\6.8\%\\1.70\%\end{tabular}     & \begin{tabular}[c]{@{}l@{}}\\3.14\\3.25\\3.15\\3.24\end{tabular}                       & \begin{tabular}[c]{@{}l@{}}\\3.82\\3.84\\3.82\\3.93\end{tabular}                         & \begin{tabular}[c]{@{}l@{}}\\79.62\%\\76.61\%\\79.15\%\\78.19\%\end{tabular}    \\
\begin{tabular}[c]{@{}l@{}}\textbf{Rating-based}~ ~\\~ ~ Non-Teenagers~\\~ ~Teenagers~~\\~ ~Non-Gender Minorities~ ~ \\~ ~Gender Minorities\end{tabular}                           & \begin{tabular}[c]{@{}l@{}}\\87.19\%\\\textbf{94.76\%}\\88.08\%\\\textbf{95.83\%}\end{tabular}    & \begin{tabular}[c]{@{}l@{}}\\9.28\%\\4.08\%\\8.65\%\\3.56\%\end{tabular}      & \begin{tabular}[c]{@{}l@{}}\\4.53\\4.78\\4.56\\4.85\end{tabular}    & \begin{tabular}[c]{@{}l@{}}\\4.6\%\\\textbf{0.47\%}\\4\%\\\textbf{1.48\%}\end{tabular}        & \begin{tabular}[c]{@{}l@{}}\\3.62\\3.38\\3.6\\3.18\end{tabular}                        & \begin{tabular}[c]{@{}l@{}}\\2.92\\2.92\\2.78\\2.78\end{tabular}                         & \begin{tabular}[c]{@{}l@{}}\\79.7\%\\83.95\%\\79.98\%\\87.74\%\end{tabular}     \\
\begin{tabular}[c]{@{}l@{}}\textbf{Blocking-based}~ ~\\~ ~ Non-Teenagers~ ~ \\~ ~Teenagers~ ~ \\~ ~Non-Gender Minorities~ ~ \\~ ~Gender Minorities\end{tabular}                    & \begin{tabular}[c]{@{}l@{}}\\76.97\%\\88.23\%\\78.34\%\\89.57\%\end{tabular}    & \begin{tabular}[c]{@{}l@{}}\\16.9\%\\9.67\%\\15.98\%\\9.48\%\end{tabular}     & \begin{tabular}[c]{@{}l@{}}\\4.15\\4.52\\4.19\\4.59\end{tabular}    & \begin{tabular}[c]{@{}l@{}}\\3.53\%\\0.31\%\\3.1\%\\0.61\%\end{tabular}     & \begin{tabular}[c]{@{}l@{}}\\3.68\\3.6\\3.67\\3.62\end{tabular}                        & \begin{tabular}[c]{@{}l@{}}\\2.75\\2.73\\2.74\\2.77\end{tabular}                         & \begin{tabular}[c]{@{}l@{}}\\82.23\%\\83.49\%\\82.47\%\\82.22\%\end{tabular}    \\
\begin{tabular}[c]{@{}l@{}}\textbf{Rating-blocking}\\\textbf{-combined}~ ~\\~ ~ Non-Teenagers~ ~ \\~ ~Teenagers~ ~ \\~ ~Non-Gender Minorities~ ~ \\~ ~Gender Minorities\end{tabular}          & \begin{tabular}[c]{@{}l@{}}\\~\\86.84\%\\\textbf{94.35\%}\\87.72\%\\\textbf{95.74\%}\end{tabular} & \begin{tabular}[c]{@{}l@{}}\\~\\9.52\%\\4.36\%\\8.9\%\\3.60\%\end{tabular}    & \begin{tabular}[c]{@{}l@{}}\\~\\4.52\\4.77\\4.55\\4.84\end{tabular} & \begin{tabular}[c]{@{}l@{}}\\~\\3.45\%\\\textbf{0.30\%}\\3\%\\\textbf{0.76\%}\end{tabular}    & \begin{tabular}[c]{@{}l@{}}\\~\\3.56\\3.31\\3.55\\3.1\end{tabular}                     & \begin{tabular}[c]{@{}l@{}}\\~\\3.07\\2.9\\3.06\\2.72\end{tabular}                       & \begin{tabular}[c]{@{}l@{}}\\~\\80.08\%\\84.86\%\\80.5\%\\87.64\%\end{tabular}  \\
\begin{tabular}[c]{@{}l@{}}\textbf{Filter-based}~ ~\\~ ~ Non-Teenagers~ ~ \\~ ~Teenagers~ ~ \\~ ~Non-Gender Minorities~ ~ \\~ ~Gender Minorities\end{tabular}                      & \begin{tabular}[c]{@{}l@{}}\\73.36\%\\89.18\%\\75.42\%\\\textbf{97.59\%}\end{tabular}    & \begin{tabular}[c]{@{}l@{}}\\20\%\\8.34\%\\18.49\%\\1.92\%\end{tabular}       & \begin{tabular}[c]{@{}l@{}}\\4.01\\4.57\\4.08\\4.91\end{tabular}    & \begin{tabular}[c]{@{}l@{}}\\7.55\%\\0.83\%\\6.46\%\\\textbf{0.05\%}\end{tabular}    & \begin{tabular}[c]{@{}l@{}}\\3.55\\2.17\\3.3\\2.33\end{tabular}                        & \begin{tabular}[c]{@{}l@{}}\\3.9\\3.23\\3.86\\3.99\end{tabular}                          & \begin{tabular}[c]{@{}l@{}}\\67.66\%\\90.22\%\\70.78\%\\84.69\%\end{tabular}    \\
\bottomrule
\end{tabular}
\end{table}

Overall, the results of our virtual experiments followed intuition in that different matching policies served different goals. First-Come-First-Serve and Blocking-based algorithms outperformed others in increasing matching rate, while the rating-based algorithm resulted (as expected) in the highest average rating. Excluding the rating-based optimized algorithms, the blocking-based algorithm also performed fairly well at an average rating of 4.21 over the replication of ground truth which had an average rating of 4.19 (out of 5 stars). We also found that optimizing for decreasing blocking rates was effective and majorly decreased the proportion of support-seeker and volunteer counselor pairs who engage in blocking from 5.3\% to 2.86\%. 

We also break down each algorithms' performance by non-teenager versus teenager versus non-gender minorities versus gender minorities groups as shown in Table \ref{Table:broken_down_simulated_result}. We note a key finding that both the rating-based and rating-blocking-combined algorithms performed extremely well among both teenager and gender minority groups. Among teenager groups, the rating-based algorithm resulted in ~95\% high chat ratings and under 0.5\% percentage of blocked pairs, while among the gender minority group, the proportion of high ratings reaches 95.83\% and the percentage of blocked pairs is 1.48\%. The rating-blocking-combined algorithm resulted in over 94\% high chat ratings with under 1\% proportion of blocked pairs for both teenagers and gender minorities. As a result, we surprisingly found that optimizing for objective metrics such as rating and blocking was able to also attain a goal of high performance among vulnerable groups.

Our agent-based simulation uncovered clear trade-offs in determining a "best" algorithm in OMHCs. For example, the filter-based method was worse than the replication of the current research site system in every metric, with the exception of extremely high performance among gender minorities; it resulted in an impressive percentage of high chat ratings at 97.59\% and extremely low percentage of blocked pairs at 0.05\%. This result follows previous work that found LGBTQ+ users in OMHCs feel safer speaking with other LGBTQ+ users \cite{Fang_2022}, but also reveals a likely exchange for lower overall satisfaction among majority groups on a platform such as the research site. We also found that all algorithms show a higher waiting time for users to be matched, as algorithmic matching in general is a trade-off with waiting time given potential delay from support-seekers waiting for a best fit volunteer counselor. The simulation also uncovered surprising results, such as that similarity-based method performed poorly on nearly all metrics. Although past work has found that people prefer to chat with those similar to themselves, it is possible that cosine similarity on demographic characteristics is not the best way to reflect these preferences. Additionally, inclusion of certain fields such as sign-up date may be less relevant to users’ interests compared to gender and age.

\section{Discussion}
\subsection{Agent-based modeling for the design of OMHCs}
In this work, we showcased our important contribution of applying agent-based simulation to the OMHC context and provided a framework for how it can help the creators and designers of online communities weigh the various trade-offs when building mechanisms to best help their support-seekers find meaningful relationships online. Our study suggests that different goals can be achieved through different algorithmic choices for these communities, from optimizing the quality of conversations to the protection of minorities on these platforms. 

It is important to note that there does not exist a universally best design for all communities; instead, the choice of algorithms and mechanisms for mental health communities is specific to the community’s context, its goals, and its support-seekers. However, although our experiments are not necessarily generalizable to every OMHC, they do provide some initial insight into making algorithmic choices for these communities. For example, communities aiming to optimize the number of users that engage in chats may wish to experiment with first-come-first-serve methods given our simulation’s results that it has the highest overall matching rate. Goals to optimize the chat experiences of users may be reached by targeting rating and blocking metrics instead. Importantly, we note that rating- and blocking-based approaches may also serve to protect vulnerable groups, as shown in Section 7.3 and discussed before. Despite these algorithms lacking explicit restrictions meant to aid these groups, such as the filter-based model, their high chat rating and blocking performances within youth and gender minority groups while using objective metrics like chat rating and blocking data indicates that strict filters are not necessarily required to serve the interests of vulnerable groups. Our stakeholders in our study’s research site consider chat ratings to be the primary indicator of user experiences, and thus find that rating-based algorithms are the most relevant and effective for their platform. 

Our research was guided by previous literature that revealed how algorithmic matching is particularly beneficial in the online mental health context, and the numerous considerations that are necessary for effective partnerships to aid users’ well-being. Through simulating the algorithms and mechanisms of the research site, we found that there are numerous trade-offs to be made in deciding how to match users together in OMHCs. For example, algorithmic matching may lead to better conversations between users but also increase the time that users must wait for a match. Our simulation revealed that there is indeed substantial improvement that can be made in bettering the quality of conversations using algorithmic matching, such as an almost 15\% increase in the number of highly rated chats given the rating-based algorithm. However, users may become impatient while waiting for a “best fit” chat partner, log off the site, and thus fewer people overall are helped by the platform. We see this reflected in our experiments - for example, first-come-first-serve is able to match the most people, with a matching rate of nearly 82\% despite being the simplest algorithm without optimization for users’ characteristics. We would also expect for there to be the lowest waiting time for first-come-first-serve given intuition, despite our experiments showing that it has the highest waiting time. However, this result may be misleading given that first-come-first-serve was able to match more support-seekers overall (i.e. highest matching rate) before support-seekers' patience levels were exceeded. Other algorithms showing a lower average waiting time among matched pairs actually had lower overall matching rates, thus having lost support-seekers who exceeded their patience level and leading to what appears to be a lower overall waiting time among matched pairs. Given this finding, we recommend that designers of algorithmic matching evaluate waiting time metrics in conjunction with the overall matching rate in order to paint a full picture of an algorithm’s trade-offs.

\subsection{Reactions of Community Stakeholders}
We also presented our results to the executives of our study's research site, including its CEO/founder and engineering lead, to gather their first impressions of our findings. In general, the the research site team found that the rating-based, blocking-based, and rating-blocking-combined methods did the best in achieving these stakeholders’ preferences for improving overall support-seeker satisfaction on the platform, and importantly did not reduce the general matching rate; as a result, support-seekers on the site would see improved experiences while chatting and the research site could continue to aid the huge majority of its large support-seeker population. Additionally, our conversations with the stakeholders of the research site considered other important factors regarding the introduction of algorithmic matching on the platform, including the computational resources and refactoring of the codebase to implement matching. It is also worth noting that the OMHC stakeholders wanted to know more about the reasoning behind the numerical results for each matching algorithm. As a result, we believe it is important that presentation of these experiments is accompanied by model explanation and algorithmic transparency as well in order to best help community stakeholders understand results and best make decisions based on these types of simulation. Addtionally, as previously mentioned, this simulation will become an open-source project, such that OMHC designers and even students can learn, explore, and understand the trade-offs in designing real-world matching algorithms for their communities.

\subsection{Limitations}
Our work has several limitations. 

First, although our paper’s primary contribution is using agent-based modeling to show how we can simulate algorithmic outcomes for online mental health communities, we note that we cannot replicate the complex systems of these communities completely. For example, we were not able to replicate the ability for volunteer counselors on the research site to pick up multiple chats if they desire or the choice of support-seekers to add a topic tag to their chat request. 

Secondly, chat ratings and blocking are just two possible outcome metrics to evaluate the performance of a support-seeker and volunteer counselor pair. These measurements have limitations; for example, volunteer counselors are unable to rate chats and support-seekers may use the 1 to 5 stars scale differently. There are other possible metrics that may have been good measurements of the performance of a support-seeker and volunteer counselor pair. For example, the research site periodically sends out emotional wellness tests to support-seekers to evaluate how their mental health has improved over time; unfortunately, during our data collection period, support-seekers rarely completed these tests and thus we did not have adequate data to analyze these results for our simulation. Retention is another metric that could be utilized to evaluate how a support-seeker and volunteer counselor pair impacted the users; however, retention is an unclear metric in the OMHC contexts as it can indicate either failure or success of the community \cite{Massimi2014-un}. 

Thirdly, the prediction models utilized in building our matching simulations were used to evaluate their outcomes, given limitations we had in being able to identify the resulting metrics of the different simulated algorithms. As a result, it is possible that the quality of matching predictions in this work are different if implemented in reality. However, we note our paper’s primary contribution of showcasing the simulation framework and to demonstrate its use in context. Lastly, we note the reaction of the research site stakeholders in that our interaction models and simulation may have been difficult to understand, and that further explanation and transparency may be needed for this work to be the most effective in real-world settings for OMHC designers.

\section{Conclusion}
In this paper, we utilized agent-based modeling in the online mental health context to reveal trade-offs of algorithmically matching peers. Evaluating data from a large and active online mental health community platform, we provided a simulation model to compare current matching mechanisms and various algorithmic matching policies, and observed their differing effects on outcome metrics including wait time and chat experiences of support-seekers. Our results indicated that algorithmic matching policies based on the applicant-proposing deferred-acceptance algorithm can lead to better chat experiences for OMHC support-seekers while still matching them for chats quickly. Our simulation can aid designers of OMHC and other online communities with a need for matching through enlightening the tensions between goals of matching as well as its impact on different communities.

\bibliographystyle{ACM-Reference-Format}
\bibliography{references}


\begin{thebibliography}{65}


\ifx \showCODEN    \undefined \def \showCODEN     #1{\unskip}     \fi
\ifx \showDOI      \undefined \def \showDOI       #1{#1}\fi
\ifx \showISBNx    \undefined \def \showISBNx     #1{\unskip}     \fi
\ifx \showISBNxiii \undefined \def \showISBNxiii  #1{\unskip}     \fi
\ifx \showISSN     \undefined \def \showISSN      #1{\unskip}     \fi
\ifx \showLCCN     \undefined \def \showLCCN      #1{\unskip}     \fi
\ifx \shownote     \undefined \def \shownote      #1{#1}          \fi
\ifx \showarticletitle \undefined \def \showarticletitle #1{#1}   \fi
\ifx \showURL      \undefined \def \showURL       {\relax}        \fi
\providecommand\bibfield[2]{#2}
\providecommand\bibinfo[2]{#2}
\providecommand\natexlab[1]{#1}
\providecommand\showeprint[2][]{arXiv:#2}

\bibitem[Abdulkadiro{\u g}lu et~al\mbox{.}(2005a)]%
        {Abdulkadiroglu2005-wk}
\bibfield{author}{\bibinfo{person}{Atila Abdulkadiro{\u g}lu},
  \bibinfo{person}{Parag~A Pathak}, {and} \bibinfo{person}{Alvin~E Roth}.}
  \bibinfo{year}{2005}\natexlab{a}.
\newblock \showarticletitle{The New York City High School Match}.
\newblock \bibinfo{journal}{\emph{Am. Econ. Rev.}} \bibinfo{volume}{95},
  \bibinfo{number}{2} (\bibinfo{date}{May} \bibinfo{year}{2005}),
  \bibinfo{pages}{364--367}.
\newblock


\bibitem[Abdulkadiro{\u g}lu et~al\mbox{.}(2005b)]%
        {Abdulkadiroglu2005-yn}
\bibfield{author}{\bibinfo{person}{Atila Abdulkadiro{\u g}lu},
  \bibinfo{person}{Parag~A Pathak}, \bibinfo{person}{Alvin~E Roth}, {and}
  \bibinfo{person}{Tayfun S{\"o}nmez}.} \bibinfo{year}{2005}\natexlab{b}.
\newblock \showarticletitle{The Boston Public School Match}.
\newblock \bibinfo{journal}{\emph{Am. Econ. Rev.}} \bibinfo{volume}{95},
  \bibinfo{number}{2} (\bibinfo{date}{May} \bibinfo{year}{2005}),
  \bibinfo{pages}{368--371}.
\newblock


\bibitem[Administration et~al\mbox{.}(2001)]%
        {Administration2001-gx}
\bibfield{author}{\bibinfo{person}{U~S Department of~Health Administration},
  \bibinfo{person}{Human Services;~Substance Abuse},
  \bibinfo{person}{Mental~Health Services}, {and} \bibinfo{person}{{US
  Department of Health and Human Services; Substance Abuse and Mental Health
  Services Administration}}.} \bibinfo{year}{2001}\natexlab{}.
\newblock \bibinfo{title}{Mental Health: Culture, Race, and Ethnicity: A
  Supplement to Mental Health A Report of the Surgeon General}.
\newblock
\newblock


\bibitem[Alvarez-Galvez(2016)]%
        {Alvarez-Galvez2016-oo}
\bibfield{author}{\bibinfo{person}{Javier Alvarez-Galvez}.}
  \bibinfo{year}{2016}\natexlab{}.
\newblock \bibinfo{title}{Network Models of Minority Opinion Spreading}.
\newblock , \bibinfo{numpages}{567-581}~pages.
\newblock


\bibitem[Alvarez-Jimenez et~al\mbox{.}(2016)]%
        {Alvarez-Jimenez2016-uz}
\bibfield{author}{\bibinfo{person}{M Alvarez-Jimenez}, \bibinfo{person}{J~F
  Gleeson}, \bibinfo{person}{S Rice}, \bibinfo{person}{C Gonzalez-Blanch},
  {and} \bibinfo{person}{S Bendall}.} \bibinfo{year}{2016}\natexlab{}.
\newblock \bibinfo{title}{Online peer-to-peer support in youth mental health:
  seizing the opportunity}.
\newblock , \bibinfo{numpages}{123--126}~pages.
\newblock


\bibitem[Andalibi et~al\mbox{.}(2021)]%
        {andalibi2021considerations}
\bibfield{author}{\bibinfo{person}{Nazanin Andalibi},
  \bibinfo{person}{Madison~K Flood}, {et~al\mbox{.}}}
  \bibinfo{year}{2021}\natexlab{}.
\newblock \showarticletitle{Considerations in designing digital peer support
  for mental health: Interview study among users of a digital support system
  (Buddy Project)}.
\newblock \bibinfo{journal}{\emph{JMIR mental health}} \bibinfo{volume}{8},
  \bibinfo{number}{1} (\bibinfo{year}{2021}), \bibinfo{pages}{e21819}.
\newblock


\bibitem[Augustaitis et~al\mbox{.}(2021)]%
        {Augustaitis2021-fm}
\bibfield{author}{\bibinfo{person}{Laima Augustaitis},
  \bibinfo{person}{Leland~A Merrill}, \bibinfo{person}{Kristi~E Gamarel}, {and}
  \bibinfo{person}{Oliver~L Haimson}.} \bibinfo{year}{2021}\natexlab{}.
\newblock \showarticletitle{Online Transgender Health Information Seeking:
  Facilitators, Barriers, and Future Directions}.
\newblock In \bibinfo{booktitle}{\emph{Proceedings of the 2021 {CHI} Conference
  on Human Factors in Computing Systems}}. \bibinfo{publisher}{Association for
  Computing Machinery}, \bibinfo{address}{New York, NY, USA},
  \bibinfo{pages}{1--14}.
\newblock


\bibitem[Baumel et~al\mbox{.}(2016)]%
        {baumel2016adaptation}
\bibfield{author}{\bibinfo{person}{Amit Baumel}, \bibinfo{person}{Christoph~U
  Correll}, {and} \bibinfo{person}{Michael Birnbaum}.}
  \bibinfo{year}{2016}\natexlab{}.
\newblock \showarticletitle{Adaptation of a peer based online emotional support
  program as an adjunct to treatment for people with schizophrenia-spectrum
  disorders}.
\newblock \bibinfo{journal}{\emph{Internet interventions}}  \bibinfo{volume}{4}
  (\bibinfo{year}{2016}), \bibinfo{pages}{35--42}.
\newblock


\bibitem[Bian and Holtzman(2011)]%
        {Bian2011-xq}
\bibfield{author}{\bibinfo{person}{Li Bian} {and} \bibinfo{person}{Henry
  Holtzman}.} \bibinfo{year}{2011}\natexlab{}.
\newblock \showarticletitle{Online friend recommendation through personality
  matching and collaborative filtering}.
\newblock \bibinfo{journal}{\emph{Proc. of UBICOMM}} \bibinfo{volume}{2011},
  \bibinfo{number}{5} (\bibinfo{year}{2011}), \bibinfo{pages}{230--235}.
\newblock


\bibitem[Birchwood and Singh(2013)]%
        {Birchwood2013-lg}
\bibfield{author}{\bibinfo{person}{Max Birchwood} {and}
  \bibinfo{person}{Swaran~P Singh}.} \bibinfo{year}{2013}\natexlab{}.
\newblock \showarticletitle{Mental health services for young people: matching
  the service to the need}.
\newblock \bibinfo{journal}{\emph{Br. J. Psychiatry Suppl.}}
  \bibinfo{volume}{54} (\bibinfo{date}{Jan.} \bibinfo{year}{2013}),
  \bibinfo{pages}{s1--2}.
\newblock


\bibitem[Bockting et~al\mbox{.}(2013)]%
        {Bockting2013-rr}
\bibfield{author}{\bibinfo{person}{Walter~O Bockting},
  \bibinfo{person}{Michael~H Miner}, \bibinfo{person}{Rebecca~E
  Swinburne~Romine}, \bibinfo{person}{Autumn Hamilton}, {and}
  \bibinfo{person}{Eli Coleman}.} \bibinfo{year}{2013}\natexlab{}.
\newblock \showarticletitle{Stigma, mental health, and resilience in an online
  sample of the {US} transgender population}.
\newblock \bibinfo{journal}{\emph{Am. J. Public Health}} \bibinfo{volume}{103},
  \bibinfo{number}{5} (\bibinfo{date}{May} \bibinfo{year}{2013}),
  \bibinfo{pages}{943--951}.
\newblock


\bibitem[Bracke et~al\mbox{.}(2008)]%
        {Bracke2008-pd}
\bibfield{author}{\bibinfo{person}{Piet Bracke}, \bibinfo{person}{Wendy
  Christiaens}, {and} \bibinfo{person}{Mieke Verhaeghe}.}
  \bibinfo{year}{2008}\natexlab{}.
\newblock \showarticletitle{Self-esteem, self-efficacy, and the balance of peer
  support among persons with chronic mental health problems: Balanced peer
  support and subjective well-being}.
\newblock \bibinfo{journal}{\emph{J. Appl. Soc. Psychol.}}
  \bibinfo{volume}{38}, \bibinfo{number}{2} (\bibinfo{date}{Jan.}
  \bibinfo{year}{2008}), \bibinfo{pages}{436--459}.
\newblock


\bibitem[Campbell and Campbell(2007)]%
        {Campbell2007-rz}
\bibfield{author}{\bibinfo{person}{Toni~A Campbell} {and}
  \bibinfo{person}{David~E Campbell}.} \bibinfo{year}{2007}\natexlab{}.
\newblock \showarticletitle{Outcomes of mentoring at‐risk college students:
  gender and ethnic matching effects}.
\newblock \bibinfo{journal}{\emph{Mentoring \& Tutoring: Partnership in
  Learning}} \bibinfo{volume}{15}, \bibinfo{number}{2} (\bibinfo{date}{May}
  \bibinfo{year}{2007}), \bibinfo{pages}{135--148}.
\newblock


\bibitem[Chawla et~al\mbox{.}(2002)]%
        {Chawla2002-hn}
\bibfield{author}{\bibinfo{person}{N~V Chawla}, \bibinfo{person}{K~W Bowyer},
  \bibinfo{person}{L~O Hall}, {and} \bibinfo{person}{W~P Kegelmeyer}.}
  \bibinfo{year}{2002}\natexlab{}.
\newblock \showarticletitle{{SMOTE}: Synthetic Minority Over-sampling
  Technique}.
\newblock \bibinfo{journal}{\emph{J. Artif. Intell. Res.}}
  \bibinfo{volume}{16} (\bibinfo{date}{June} \bibinfo{year}{2002}),
  \bibinfo{pages}{321--357}.
\newblock


\bibitem[Cipolletta et~al\mbox{.}(2017)]%
        {Cipolletta2017-by}
\bibfield{author}{\bibinfo{person}{Sabrina Cipolletta},
  \bibinfo{person}{Riccardo Votadoro}, {and} \bibinfo{person}{Elena Faccio}.}
  \bibinfo{year}{2017}\natexlab{}.
\newblock \showarticletitle{Online support for transgender people: an analysis
  of forums and social networks}.
\newblock \bibinfo{journal}{\emph{Health Soc. Care Community}}
  \bibinfo{volume}{25}, \bibinfo{number}{5} (\bibinfo{date}{Sept.}
  \bibinfo{year}{2017}), \bibinfo{pages}{1542--1551}.
\newblock


\bibitem[Coleman et~al\mbox{.}(1995)]%
        {Coleman_undated-xf}
\bibfield{author}{\bibinfo{person}{Hardin L~K Coleman},
  \bibinfo{person}{Bruce~E Wampold}, {and} \bibinfo{person}{Sherry~L Casali}.}
  \bibinfo{year}{1995}\natexlab{}.
\newblock \showarticletitle{Ethnic minorities' ratings of ethnically similar
  and European American counselors: A meta-analysis}.
\newblock \bibinfo{journal}{\emph{J. Couns. Psychol.}} \bibinfo{volume}{42},
  \bibinfo{number}{1} (\bibinfo{date}{Jan.} \bibinfo{year}{1995}),
  \bibinfo{pages}{55--64}.
\newblock


\bibitem[Craig and McInroy(2014)]%
        {Craig2014-yl}
\bibfield{author}{\bibinfo{person}{Shelley~L Craig} {and}
  \bibinfo{person}{Lauren McInroy}.} \bibinfo{year}{2014}\natexlab{}.
\newblock \showarticletitle{You Can Form a Part of Yourself Online: The
  Influence of New Media on Identity Development and Coming Out for {LGBTQ}
  Youth}.
\newblock \bibinfo{journal}{\emph{J. Gay Lesbian Ment. Health}}
  \bibinfo{volume}{18}, \bibinfo{number}{1} (\bibinfo{date}{Jan.}
  \bibinfo{year}{2014}), \bibinfo{pages}{95--109}.
\newblock


\bibitem[De~Choudhury and De(2014)]%
        {De_Choudhury2014-yc}
\bibfield{author}{\bibinfo{person}{Munmun De~Choudhury} {and}
  \bibinfo{person}{Sushovan De}.} \bibinfo{year}{2014}\natexlab{}.
\newblock \showarticletitle{Mental Health Discourse on reddit:
  {Self-Disclosure}, Social Support, and Anonymity}.
\newblock \bibinfo{journal}{\emph{Eighth International {AAAI} Conference on
  Weblogs and Social Media}}  \bibinfo{volume}{2014} (\bibinfo{date}{May}
  \bibinfo{year}{2014}), \bibinfo{pages}{71--80}.
\newblock


\bibitem[De~Choudhury and K{\i}c{\i}man(2017)]%
        {De_Choudhury2017-ox}
\bibfield{author}{\bibinfo{person}{Munmun De~Choudhury} {and}
  \bibinfo{person}{Emre K{\i}c{\i}man}.} \bibinfo{year}{2017}\natexlab{}.
\newblock \showarticletitle{The Language of Social Support in Social Media and
  its Effect on Suicidal Ideation Risk}.
\newblock \bibinfo{journal}{\emph{Proc Int AAAI Conf Weblogs Soc Media}}
  \bibinfo{volume}{2017} (\bibinfo{date}{May} \bibinfo{year}{2017}),
  \bibinfo{pages}{32--41}.
\newblock


\bibitem[Delcea et~al\mbox{.}(2017)]%
        {Delcea2017-vp}
\bibfield{author}{\bibinfo{person}{Camelia Delcea},
  \bibinfo{person}{Ioana~Alexandra Bradea}, \bibinfo{person}{Liviu~Adrian
  Cotfas}, {and} \bibinfo{person}{Emil Scarlat}.}
  \bibinfo{year}{2017}\natexlab{}.
\newblock \bibinfo{title}{Opinion influence in online social media environments
  --- {U} grey system theory and agent-based modeling approach}.
\newblock
\newblock


\bibitem[Doherty et~al\mbox{.}(2012)]%
        {doherty2012engagement}
\bibfield{author}{\bibinfo{person}{Gavin Doherty}, \bibinfo{person}{David
  Coyle}, {and} \bibinfo{person}{John Sharry}.}
  \bibinfo{year}{2012}\natexlab{}.
\newblock \showarticletitle{Engagement with online mental health interventions:
  an exploratory clinical study of a treatment for depression}. In
  \bibinfo{booktitle}{\emph{Proceedings of the SIGCHI Conference on Human
  Factors in Computing Systems}}. \bibinfo{pages}{1421--1430}.
\newblock


\bibitem[Du et~al\mbox{.}(2017)]%
        {Du2017-jm}
\bibfield{author}{\bibinfo{person}{Erhu Du}, \bibinfo{person}{Ximing Cai},
  \bibinfo{person}{Zhiyong Sun}, {and} \bibinfo{person}{Barbara Minsker}.}
  \bibinfo{year}{2017}\natexlab{}.
\newblock \showarticletitle{Exploring the role of social media and individual
  behaviors in flood evacuation processes: An agent-based modeling approach}.
\newblock \bibinfo{journal}{\emph{Water Resour. Res.}} \bibinfo{volume}{53},
  \bibinfo{number}{11} (\bibinfo{date}{Nov.} \bibinfo{year}{2017}),
  \bibinfo{pages}{9164--9180}.
\newblock


\bibitem[Fan et~al\mbox{.}(2014)]%
        {Fan2014-db}
\bibfield{author}{\bibinfo{person}{Hanmei Fan}, \bibinfo{person}{Reeva
  Lederman}, \bibinfo{person}{Stephen~P Smith}, {and} \bibinfo{person}{Shanton
  Chang}.} \bibinfo{year}{2014}\natexlab{}.
\newblock \showarticletitle{How Trust Is Formed in Online Health Communities: A
  Process Perspective}.
\newblock \bibinfo{journal}{\emph{Communications of the Association for
  Information Systems}} \bibinfo{volume}{34}, \bibinfo{number}{1}
  (\bibinfo{year}{2014}), \bibinfo{pages}{28}.
\newblock


\bibitem[Fang and Zhu(2022)]%
        {Fang_2022}
\bibfield{author}{\bibinfo{person}{Anna Fang} {and} \bibinfo{person}{Haiyi
  Zhu}.} \bibinfo{year}{2022}\natexlab{}.
\newblock \showarticletitle{Matching for Peer Support: Exploring Algorithmic
  Matching for Online Mental Health Communities}.
\newblock \bibinfo{journal}{\emph{Proceedings of the 25th {ACM} Conference On
  Computer-Supported Cooperative Work And Social Computing}}
  (\bibinfo{year}{2022}).
\newblock


\bibitem[Felton(1986)]%
        {Felton1986-kf}
\bibfield{author}{\bibinfo{person}{Judith~R Felton}.}
  \bibinfo{year}{1986}\natexlab{}.
\newblock \showarticletitle{Sex makes a difference---How gender affects the
  therapeutic relationship}.
\newblock \bibinfo{journal}{\emph{Clin. Soc. Work J.}} \bibinfo{volume}{14},
  \bibinfo{number}{2} (\bibinfo{date}{June} \bibinfo{year}{1986}),
  \bibinfo{pages}{127--138}.
\newblock


\bibitem[Fenwick and Neal(2001)]%
        {Fenwick2001-qm}
\bibfield{author}{\bibinfo{person}{Graham~D Fenwick} {and}
  \bibinfo{person}{Derrick~J Neal}.} \bibinfo{year}{2001}\natexlab{}.
\newblock \showarticletitle{Effect of gender composition on group performance}.
\newblock \bibinfo{journal}{\emph{Gend. Work Organ.}} \bibinfo{volume}{8},
  \bibinfo{number}{2} (\bibinfo{date}{April} \bibinfo{year}{2001}),
  \bibinfo{pages}{205--225}.
\newblock


\bibitem[Fish et~al\mbox{.}(2020)]%
        {Fish2020-by}
\bibfield{author}{\bibinfo{person}{Jessica~N Fish}, \bibinfo{person}{Lauren~B
  McInroy}, \bibinfo{person}{Megan~S Paceley}, \bibinfo{person}{Natasha~D
  Williams}, \bibinfo{person}{Sara Henderson}, \bibinfo{person}{Deborah~S
  Levine}, {and} \bibinfo{person}{Rachel~N Edsall}.}
  \bibinfo{year}{2020}\natexlab{}.
\newblock \showarticletitle{``I'm Kinda Stuck at Home With Unsupportive Parents
  Right Now'': {LGBTQ} Youths' Experiences With {COVID-19} and the Importance
  of Online Support}.
\newblock \bibinfo{journal}{\emph{J. Adolesc. Health Care}}
  \bibinfo{volume}{67}, \bibinfo{number}{3} (\bibinfo{date}{Sept.}
  \bibinfo{year}{2020}), \bibinfo{pages}{450--452}.
\newblock


\bibitem[Flaskerud and Liu(1991)]%
        {Flaskerud1991-wo}
\bibfield{author}{\bibinfo{person}{J~H Flaskerud} {and} \bibinfo{person}{P~Y
  Liu}.} \bibinfo{year}{1991}\natexlab{}.
\newblock \showarticletitle{Effects of an Asian client-therapist language,
  ethnicity and gender match on utilization and outcome of therapy}.
\newblock \bibinfo{journal}{\emph{Community Ment. Health J.}}
  \bibinfo{volume}{27}, \bibinfo{number}{1} (\bibinfo{date}{Feb.}
  \bibinfo{year}{1991}), \bibinfo{pages}{31--42}.
\newblock


\bibitem[Gale and Shapley(1961)]%
        {Gale1961-ig}
\bibfield{author}{\bibinfo{person}{D Gale} {and} \bibinfo{person}{L~S
  Shapley}.} \bibinfo{year}{1961}\natexlab{}.
\newblock \bibinfo{title}{{COLLEGE} {ADMISSIONS} {AND} {THE} {STABILITY} {OF}
  {MARRIAGE}}.
\newblock
\newblock


\bibitem[Graham et~al\mbox{.}(2014)]%
        {Graham2014-ci}
\bibfield{author}{\bibinfo{person}{Louis~F Graham}, \bibinfo{person}{Halley~P
  Crissman}, \bibinfo{person}{Jack Tocco}, \bibinfo{person}{Laura~A Hughes},
  \bibinfo{person}{Rachel~C Snow}, {and} \bibinfo{person}{Mark~B Padilla}.}
  \bibinfo{year}{2014}\natexlab{}.
\newblock \showarticletitle{Interpersonal Relationships and Social Support in
  Transitioning Narratives of Black Transgender Women in Detroit}.
\newblock \bibinfo{journal}{\emph{International Journal of Transgenderism}}
  \bibinfo{volume}{15}, \bibinfo{number}{2} (\bibinfo{date}{April}
  \bibinfo{year}{2014}), \bibinfo{pages}{100--113}.
\newblock


\bibitem[Gui et~al\mbox{.}(2017)]%
        {gui2017investigating}
\bibfield{author}{\bibinfo{person}{Xinning Gui}, \bibinfo{person}{Yu Chen},
  \bibinfo{person}{Yubo Kou}, \bibinfo{person}{Katie Pine}, {and}
  \bibinfo{person}{Yunan Chen}.} \bibinfo{year}{2017}\natexlab{}.
\newblock \showarticletitle{Investigating support seeking from peers for
  pregnancy in online health communities}.
\newblock \bibinfo{journal}{\emph{Proceedings of the ACM on Human-Computer
  Interaction}} \bibinfo{volume}{1}, \bibinfo{number}{CSCW}
  (\bibinfo{year}{2017}), \bibinfo{pages}{1--19}.
\newblock


\bibitem[Hitsch et~al\mbox{.}(2010)]%
        {Hitsch2010-nb}
\bibfield{author}{\bibinfo{person}{G{\"u}nter~J Hitsch}, \bibinfo{person}{Ali
  Horta{\c c}su}, {and} \bibinfo{person}{Dan Ariely}.}
  \bibinfo{year}{2010}\natexlab{}.
\newblock \showarticletitle{What makes you click?---Mate preferences in online
  dating}.
\newblock \bibinfo{journal}{\emph{Quantitative Marketing and Economics}}
  \bibinfo{volume}{8}, \bibinfo{number}{4} (\bibinfo{date}{Dec.}
  \bibinfo{year}{2010}), \bibinfo{pages}{393--427}.
\newblock


\bibitem[Huang et~al\mbox{.}(2014)]%
        {Huang2014-qj}
\bibfield{author}{\bibinfo{person}{Qingxu Huang}, \bibinfo{person}{Dawn~C
  Parker}, \bibinfo{person}{Tatiana Filatova}, {and} \bibinfo{person}{Shipeng
  Sun}.} \bibinfo{year}{2014}\natexlab{}.
\newblock \showarticletitle{A Review of Urban Residential Choice Models Using
  {Agent-Based} Modeling}.
\newblock \bibinfo{journal}{\emph{Environ. Plann. B Plann. Des.}}
  \bibinfo{volume}{41}, \bibinfo{number}{4} (\bibinfo{date}{Aug.}
  \bibinfo{year}{2014}), \bibinfo{pages}{661--689}.
\newblock


\bibitem[Jackson and Kirschner(1973)]%
        {Jackson1973-wo}
\bibfield{author}{\bibinfo{person}{Gerald~G Jackson} {and}
  \bibinfo{person}{Samuel~A Kirschner}.} \bibinfo{year}{1973}\natexlab{}.
\newblock \showarticletitle{Racial self-designation and preference for a
  counselor}.
\newblock \bibinfo{journal}{\emph{J. Couns. Psychol.}} \bibinfo{volume}{20},
  \bibinfo{number}{6} (\bibinfo{date}{Nov.} \bibinfo{year}{1973}),
  \bibinfo{pages}{560--564}.
\newblock


\bibitem[Kim et~al\mbox{.}(2008)]%
        {Kim2008-iq}
\bibfield{author}{\bibinfo{person}{Heejung~S Kim}, \bibinfo{person}{David~K
  Sherman}, {and} \bibinfo{person}{Shelley~E Taylor}.}
  \bibinfo{year}{2008}\natexlab{}.
\newblock \showarticletitle{Culture and social support}.
\newblock \bibinfo{journal}{\emph{Am. Psychol.}} \bibinfo{volume}{63},
  \bibinfo{number}{6} (\bibinfo{date}{Sept.} \bibinfo{year}{2008}),
  \bibinfo{pages}{518--526}.
\newblock


\bibitem[Leong(1986)]%
        {Leong1986-iz}
\bibfield{author}{\bibinfo{person}{Frederick~T Leong}.}
  \bibinfo{year}{1986}\natexlab{}.
\newblock \showarticletitle{Counseling and psychotherapy with
  {Asian-Americans}: Review of the literature}.
\newblock \bibinfo{journal}{\emph{J. Couns. Psychol.}} \bibinfo{volume}{33},
  \bibinfo{number}{2} (\bibinfo{date}{April} \bibinfo{year}{1986}),
  \bibinfo{pages}{196--206}.
\newblock


\bibitem[Maramba and Hall(2002)]%
        {Maramba2002-ug}
\bibfield{author}{\bibinfo{person}{Gloria~Gia Maramba} {and}
  \bibinfo{person}{Gordon C~Nagayama Hall}.} \bibinfo{year}{2002}\natexlab{}.
\newblock \showarticletitle{Meta-analyses of ethnic match as a predictor of
  dropout, utilization, and level of functioning}.
\newblock \bibinfo{journal}{\emph{Cultur. Divers. Ethnic Minor. Psychol.}}
  \bibinfo{volume}{8}, \bibinfo{number}{3} (\bibinfo{date}{Aug.}
  \bibinfo{year}{2002}), \bibinfo{pages}{290--297}.
\newblock


\bibitem[Massimi et~al\mbox{.}(2014)]%
        {Massimi2014-un}
\bibfield{author}{\bibinfo{person}{Michael Massimi}, \bibinfo{person}{Jackie~L
  Bender}, \bibinfo{person}{Holly~O Witteman}, {and} \bibinfo{person}{Osman~H
  Ahmed}.} \bibinfo{year}{2014}\natexlab{}.
\newblock \showarticletitle{Life transitions and online health communities:
  reflecting on adoption, use, and disengagement}. In
  \bibinfo{booktitle}{\emph{Proceedings of the 17th {ACM} conference on
  Computer supported cooperative work \& social computing}} (Baltimore,
  Maryland, USA) \emph{(\bibinfo{series}{CSCW '14})}.
  \bibinfo{publisher}{Association for Computing Machinery},
  \bibinfo{address}{New York, NY, USA}, \bibinfo{pages}{1491--1501}.
\newblock


\bibitem[Maultsby(1982)]%
        {Maultsby1982-nm}
\bibfield{author}{\bibinfo{person}{Maxie~C Maultsby}.}
  \bibinfo{year}{1982}\natexlab{}.
\newblock \bibinfo{title}{A historical view of Blacks' distrust of psychiatry}.
\newblock , \bibinfo{numpages}{39--55}~pages.
\newblock


\bibitem[McConnell et~al\mbox{.}(2017)]%
        {McConnell2017-tt}
\bibfield{author}{\bibinfo{person}{Elizabeth~A McConnell},
  \bibinfo{person}{Antonia Clifford}, \bibinfo{person}{Aaron~K Korpak},
  \bibinfo{person}{Gregory Phillips, 2nd}, {and} \bibinfo{person}{Michelle
  Birkett}.} \bibinfo{year}{2017}\natexlab{}.
\newblock \showarticletitle{Identity, Victimization, and Support: Facebook
  Experiences and Mental Health Among {LGBTQ} Youth}.
\newblock \bibinfo{journal}{\emph{Comput. Human Behav.}}  \bibinfo{volume}{76}
  (\bibinfo{date}{Nov.} \bibinfo{year}{2017}), \bibinfo{pages}{237--244}.
\newblock


\bibitem[Mitrovi{\'c} and Tadi{\'c}(2012)]%
        {Mitrovic2012-hh}
\bibfield{author}{\bibinfo{person}{Marija Mitrovi{\'c}} {and}
  \bibinfo{person}{Bosiljka Tadi{\'c}}.} \bibinfo{year}{2012}\natexlab{}.
\newblock \showarticletitle{Dynamics of bloggers' communities: Bipartite
  networks from empirical data and agent-based modeling}.
\newblock \bibinfo{journal}{\emph{Physica A: Statistical Mechanics and its
  Applications}} \bibinfo{volume}{391}, \bibinfo{number}{21}
  (\bibinfo{year}{2012}), \bibinfo{pages}{5264--5278}.
\newblock


\bibitem[Naslund et~al\mbox{.}(2016)]%
        {Naslund2016-gm}
\bibfield{author}{\bibinfo{person}{J~A Naslund}, \bibinfo{person}{K~A
  Aschbrenner}, \bibinfo{person}{L~A Marsch}, {and} \bibinfo{person}{S~J
  Bartels}.} \bibinfo{year}{2016}\natexlab{}.
\newblock \bibinfo{title}{The future of mental health care: peer-to-peer
  support and social media}.
\newblock , \bibinfo{numpages}{113--122}~pages.
\newblock


\bibitem[Nigel~Gilbert and {Troitzsch}(2006)]%
        {Nigel_Gilbert2006-dx}
\bibfield{author}{\bibinfo{person}{G Nigel~Gilbert} {and}
  \bibinfo{person}{{Troitzsch}}.} \bibinfo{year}{2006}\natexlab{}.
\newblock \bibinfo{booktitle}{\emph{Simulation for the Social Scientist:
  Special Reprint}}.
\newblock \bibinfo{publisher}{McGraw-Hill Education}, \bibinfo{address}{New
  York, NY}.
\newblock


\bibitem[Plikynas et~al\mbox{.}(2015)]%
        {Plikynas2015-gd}
\bibfield{author}{\bibinfo{person}{D Plikynas}, \bibinfo{person}{A Raudys},
  {and} \bibinfo{person}{S Raudys}.} \bibinfo{year}{2015}\natexlab{}.
\newblock \showarticletitle{Agent-based modelling of excitation propagation in
  social media groups}.
\newblock \bibinfo{journal}{\emph{J. Exp. Theor. Artif. Intell.}}
  \bibinfo{volume}{27}, \bibinfo{number}{4} (\bibinfo{date}{July}
  \bibinfo{year}{2015}), \bibinfo{pages}{373--388}.
\newblock


\bibitem[Prescott et~al\mbox{.}(2020a)]%
        {Prescott2020-kg}
\bibfield{author}{\bibinfo{person}{Julie Prescott},
  \bibinfo{person}{Rathbone~Amy Leigh}, {and} \bibinfo{person}{Terry Hanley}.}
  \bibinfo{year}{2020}\natexlab{a}.
\newblock \showarticletitle{Online mental health communities, self-efficacy and
  transition to further support}.
\newblock \bibinfo{journal}{\emph{Mental Health Review Journal}}
  \bibinfo{volume}{25}, \bibinfo{number}{4} (\bibinfo{date}{Jan.}
  \bibinfo{year}{2020}), \bibinfo{pages}{329--344}.
\newblock


\bibitem[Prescott et~al\mbox{.}(2020b)]%
        {Prescott2020-wy}
\bibfield{author}{\bibinfo{person}{Julie Prescott}, \bibinfo{person}{Amy~Leigh
  Rathbone}, {and} \bibinfo{person}{Gill Brown}.}
  \bibinfo{year}{2020}\natexlab{b}.
\newblock \showarticletitle{Online peer to peer support: Qualitative analysis
  of {UK} and {US} open mental health Facebook groups}.
\newblock \bibinfo{journal}{\emph{Digit Health}}  \bibinfo{volume}{6}
  (\bibinfo{date}{Jan.} \bibinfo{year}{2020}),
  \bibinfo{pages}{2055207620979209}.
\newblock


\bibitem[Ren and Kraut(2010)]%
        {Ren2010-jr}
\bibfield{author}{\bibinfo{person}{Yuqing Ren} {and} \bibinfo{person}{Robert~E
  Kraut}.} \bibinfo{year}{2010}\natexlab{}.
\newblock \showarticletitle{Agent-based modeling to inform online community
  theory and design: Impact of discussion moderation on member commitment and
  contribution}.
\newblock \bibinfo{journal}{\emph{Second round revise and resubmit at
  Information Systems Research}} \bibinfo{volume}{21}, \bibinfo{number}{3}
  (\bibinfo{year}{2010}).
\newblock


\bibitem[Ren and Kraut(2014a)]%
        {Ren2014-wb}
\bibfield{author}{\bibinfo{person}{Yuqing Ren} {and} \bibinfo{person}{Robert~E
  Kraut}.} \bibinfo{year}{2014}\natexlab{a}.
\newblock \showarticletitle{{Agent-Based} Modeling to Inform Online Community
  Design: Impact of Topical Breadth, Message Volume, and Discussion Moderation
  on Member Commitment and Contribution}.
\newblock \bibinfo{journal}{\emph{Human--Computer Interaction}}
  \bibinfo{volume}{29}, \bibinfo{number}{4} (\bibinfo{date}{July}
  \bibinfo{year}{2014}), \bibinfo{pages}{351--389}.
\newblock


\bibitem[Ren and Kraut(2014b)]%
        {Ren2014-pn}
\bibfield{author}{\bibinfo{person}{Yuqing Ren} {and} \bibinfo{person}{Robert~E
  Kraut}.} \bibinfo{year}{2014}\natexlab{b}.
\newblock \showarticletitle{Agent Based Modeling to Inform the Design of
  Multiuser Systems}.
\newblock In \bibinfo{booktitle}{\emph{Ways of Knowing in {HCI}}},
  \bibfield{editor}{\bibinfo{person}{Judith~S Olson} {and}
  \bibinfo{person}{Wendy~A Kellogg}} (Eds.). \bibinfo{publisher}{Springer New
  York}, \bibinfo{address}{New York, NY}, \bibinfo{pages}{395--419}.
\newblock


\bibitem[Roth and Peranson(1999)]%
        {Roth1999-zc}
\bibfield{author}{\bibinfo{person}{Alvin~E Roth} {and} \bibinfo{person}{Elliott
  Peranson}.} \bibinfo{year}{1999}\natexlab{}.
\newblock \showarticletitle{The Redesign of the Matching Market for American
  Physicians: Some Engineering Aspects of Economic Design}.
\newblock \bibinfo{journal}{\emph{Am. Econ. Rev.}} \bibinfo{volume}{89},
  \bibinfo{number}{4} (\bibinfo{date}{Sept.} \bibinfo{year}{1999}),
  \bibinfo{pages}{748--780}.
\newblock


\bibitem[Salzer et~al\mbox{.}(2013)]%
        {Salzer2013-gg}
\bibfield{author}{\bibinfo{person}{Mark~S Salzer}, \bibinfo{person}{Nicole
  Darr}, \bibinfo{person}{Gina Calhoun}, \bibinfo{person}{William Boyer},
  \bibinfo{person}{Randall~E Loss}, \bibinfo{person}{Jerry Goessel},
  \bibinfo{person}{Edward Schwenk}, {and} \bibinfo{person}{Eugene
  Brusilovskiy}.} \bibinfo{year}{2013}\natexlab{}.
\newblock \showarticletitle{Benefits of working as a certified peer specialist:
  results from a statewide survey}.
\newblock \bibinfo{journal}{\emph{Psychiatr. Rehabil. J.}}
  \bibinfo{volume}{36}, \bibinfo{number}{3} (\bibinfo{date}{Sept.}
  \bibinfo{year}{2013}), \bibinfo{pages}{219--221}.
\newblock


\bibitem[Schweitzer and Garcia(2010)]%
        {Schweitzer2010-pe}
\bibfield{author}{\bibinfo{person}{Frank Schweitzer} {and}
  \bibinfo{person}{David Garcia}.} \bibinfo{year}{2010}\natexlab{}.
\newblock \showarticletitle{An agent-based model of collective emotions in
  online communities}.
\newblock \bibinfo{journal}{\emph{The European Physical Journal B}}
  \bibinfo{volume}{77}, \bibinfo{number}{4} (\bibinfo{date}{June}
  \bibinfo{year}{2010}), \bibinfo{pages}{533--545}.
\newblock
\showeprint[arxiv]{1006.5305}~[physics.soc-ph]


\bibitem[Sibley and Crooks(2020)]%
        {Sibley2020-cc}
\bibfield{author}{\bibinfo{person}{Ciara Sibley} {and}
  \bibinfo{person}{Andrew~T Crooks}.} \bibinfo{year}{2020}\natexlab{}.
\newblock \showarticletitle{Exploring the effects of link recommendations on
  social networks: an agent-based modeling approach}. In
  \bibinfo{booktitle}{\emph{Proceedings of the 2020 Spring Simulation
  Conference}} (Fairfax, Virginia) \emph{(\bibinfo{series}{SpringSim '20},
  \bibinfo{number}{Article 41})}. \bibinfo{publisher}{Society for Computer
  Simulation International}, \bibinfo{address}{San Diego, CA, USA},
  \bibinfo{pages}{1--12}.
\newblock


\bibitem[Stewart and D'Mello(2018)]%
        {Stewart2018-xy}
\bibfield{author}{\bibinfo{person}{Angela Stewart} {and}
  \bibinfo{person}{Sidney~K D'Mello}.} \bibinfo{year}{2018}\natexlab{}.
\newblock \showarticletitle{Connecting the Dots Towards Collaborative {AIED}:
  Linking Group Makeup to Process to Learning}. In
  \bibinfo{booktitle}{\emph{Artificial Intelligence in Education}}.
  \bibinfo{publisher}{Springer International Publishing},
  \bibinfo{address}{Boulder, CO}, \bibinfo{pages}{545--556}.
\newblock


\bibitem[Stratford et~al\mbox{.}(2019)]%
        {Stratford2019-nq}
\bibfield{author}{\bibinfo{person}{Anthony~C Stratford}, \bibinfo{person}{Matt
  Halpin}, \bibinfo{person}{Keely Phillips}, \bibinfo{person}{Frances
  Skerritt}, \bibinfo{person}{Anne Beales}, \bibinfo{person}{Vincent Cheng},
  \bibinfo{person}{Magdel Hammond}, \bibinfo{person}{Mary O'Hagan},
  \bibinfo{person}{Catherine Loreto}, \bibinfo{person}{Kim Tiengtom},
  \bibinfo{person}{Benon Kobe}, \bibinfo{person}{Steve Harrington},
  \bibinfo{person}{Dan Fisher}, {and} \bibinfo{person}{Larry Davidson}.}
  \bibinfo{year}{2019}\natexlab{}.
\newblock \showarticletitle{The growth of peer support: an international
  charter}.
\newblock \bibinfo{journal}{\emph{J. Ment. Health}} \bibinfo{volume}{28},
  \bibinfo{number}{6} (\bibinfo{date}{Dec.} \bibinfo{year}{2019}),
  \bibinfo{pages}{627--632}.
\newblock


\bibitem[Sue(1988)]%
        {Sue1988-nz}
\bibfield{author}{\bibinfo{person}{S Sue}.} \bibinfo{year}{1988}\natexlab{}.
\newblock \showarticletitle{Psychotherapeutic services for ethnic minorities.
  Two decades of research findings}.
\newblock \bibinfo{journal}{\emph{Am. Psychol.}} \bibinfo{volume}{43},
  \bibinfo{number}{4} (\bibinfo{date}{April} \bibinfo{year}{1988}),
  \bibinfo{pages}{301--308}.
\newblock


\bibitem[Tan et~al\mbox{.}(2011)]%
        {Tan2011-bm}
\bibfield{author}{\bibinfo{person}{Zhangwen Tan}, \bibinfo{person}{Xiaochen
  Li}, {and} \bibinfo{person}{Wenji Mao}.} \bibinfo{year}{2011}\natexlab{}.
\newblock \showarticletitle{Agent-Based Modeling of Netizen Groups in Chinese
  Internet Events}.
\newblock \bibinfo{journal}{\emph{Intelligence and Security Informatics}}
  \bibinfo{volume}{2011}, \bibinfo{number}{1} (\bibinfo{year}{2011}),
  \bibinfo{pages}{43--53}.
\newblock


\bibitem[Tesfatsion and Judd(2006)]%
        {Tesfatsion2006-ag}
\bibfield{author}{\bibinfo{person}{Leigh Tesfatsion} {and}
  \bibinfo{person}{Kenneth~L Judd}.} \bibinfo{year}{2006}\natexlab{}.
\newblock \bibinfo{booktitle}{\emph{Handbook of Computational Economics:
  {Agent-Based} Computational Economics}}.
\newblock \bibinfo{publisher}{Elsevier}, \bibinfo{address}{Amsterdam, The
  Netherlands}.
\newblock


\bibitem[Turner(1999)]%
        {Turner1999-tj}
\bibfield{author}{\bibinfo{person}{G Turner}.} \bibinfo{year}{1999}\natexlab{}.
\newblock \showarticletitle{Peer support and young people's health}.
\newblock \bibinfo{journal}{\emph{J. Adolesc.}} \bibinfo{volume}{22},
  \bibinfo{number}{4} (\bibinfo{date}{Aug.} \bibinfo{year}{1999}),
  \bibinfo{pages}{567--572}.
\newblock


\bibitem[van Maanen and van~der Vecht(2013)]%
        {Van_Maanen2013-yt}
\bibfield{author}{\bibinfo{person}{Peter-Paul van Maanen} {and}
  \bibinfo{person}{Bob van~der Vecht}.} \bibinfo{year}{2013}\natexlab{}.
\newblock \bibinfo{title}{An agent-based approach to modeling online social
  influence}.
\newblock
\newblock


\bibitem[Vlahovic et~al\mbox{.}(2014)]%
        {Vlahovic2014-rp}
\bibfield{author}{\bibinfo{person}{Tatiana~A Vlahovic},
  \bibinfo{person}{Yi-Chia Wang}, \bibinfo{person}{Robert~E Kraut}, {and}
  \bibinfo{person}{John~M Levine}.} \bibinfo{year}{2014}\natexlab{}.
\newblock \showarticletitle{Support matching and satisfaction in an online
  breast cancer support community}. In \bibinfo{booktitle}{\emph{Proceedings of
  the {SIGCHI} Conference on Human Factors in Computing Systems}} (Toronto,
  Ontario, Canada) \emph{(\bibinfo{series}{CHI '14})}.
  \bibinfo{publisher}{Association for Computing Machinery},
  \bibinfo{address}{New York, NY, USA}, \bibinfo{pages}{1625--1634}.
\newblock


\bibitem[Wang et~al\mbox{.}(2015)]%
        {Wang2015-th}
\bibfield{author}{\bibinfo{person}{Z Wang}, \bibinfo{person}{J Liao},
  \bibinfo{person}{Q Cao}, \bibinfo{person}{H Qi}, {and} \bibinfo{person}{Z
  Wang}.} \bibinfo{year}{2015}\natexlab{}.
\newblock \showarticletitle{Friendbook: A {Semantic-Based} Friend
  Recommendation System for Social Networks}.
\newblock \bibinfo{journal}{\emph{IEEE Trans. Mob. Comput.}}
  \bibinfo{volume}{14}, \bibinfo{number}{3} (\bibinfo{date}{March}
  \bibinfo{year}{2015}), \bibinfo{pages}{538--551}.
\newblock


\bibitem[Whaley(2001)]%
        {Whaley2001-yi}
\bibfield{author}{\bibinfo{person}{Arthur~L Whaley}.}
  \bibinfo{year}{2001}\natexlab{}.
\newblock \showarticletitle{Cultural Mistrust and Mental Health Services for
  African Americans: A Review and {Meta-Analysis}}.
\newblock \bibinfo{journal}{\emph{Couns. Psychol.}} \bibinfo{volume}{29},
  \bibinfo{number}{4} (\bibinfo{date}{July} \bibinfo{year}{2001}),
  \bibinfo{pages}{513--531}.
\newblock


\bibitem[Whittaker et~al\mbox{.}(1997)]%
        {Whittaker1997-jd}
\bibfield{author}{\bibinfo{person}{Steve Whittaker}, \bibinfo{person}{Ellen
  Isaacs}, {and} \bibinfo{person}{Vicki O'Day}.}
  \bibinfo{year}{1997}\natexlab{}.
\newblock \bibinfo{title}{Widening the net}.
\newblock , \bibinfo{numpages}{27--30}~pages.
\newblock


\bibitem[Yadavaia and Hayes(2012)]%
        {Yadavaia2012-yg}
\bibfield{author}{\bibinfo{person}{James~E Yadavaia} {and}
  \bibinfo{person}{Steven~C Hayes}.} \bibinfo{year}{2012}\natexlab{}.
\newblock \showarticletitle{Acceptance and Commitment Therapy for {Self-Stigma}
  Around Sexual Orientation: A Multiple Baseline Evaluation}.
\newblock \bibinfo{journal}{\emph{Cogn. Behav. Pract.}} \bibinfo{volume}{19},
  \bibinfo{number}{4} (\bibinfo{date}{Nov.} \bibinfo{year}{2012}),
  \bibinfo{pages}{545--559}.
\newblock


\end{thebibliography}

\newpage
\appendix

\section{Pseudocode for Applicant-Proposing Deferred-Acceptance Algorithm}
\begin{lstlisting}

Parameter S {support-seekers available to be matched in the current simulation period}

Parameter V {volunteer counselors available to be matched in the current simulation period}

function stableMatching {
    Initialize all s in S and v in V to be available
    while there exists available s who still has a volunteer v to apply to {
       v = first volunteer counselor on support seeker s' list to whom v has not yet applied
       if v is available
         (s, v) become matched
       else some pair (s', v) already exists
         if v prefers s to s'
            s' becomes available
           (s, v) become matched 
         else
           (s', v) remain matched
    }
}
\end{lstlisting}

\end{document}